\documentclass[acmtog, preprint]{acmart}
\acmSubmissionID{TOG-24-0162}

\usepackage{booktabs} %
\usepackage{bm}
\usepackage{subcaption}
\usepackage{enumitem}

\usepackage[ruled]{algorithm2e} %

\SetAlFnt{\small}
\SetAlCapFnt{\small}
\SetAlCapNameFnt{\small}
\SetAlCapHSkip{0pt}
\SetKwInput{KwParameters}{Parameters}

\acmJournal{TOG}
\acmVolume{44}
\acmNumber{2}
\acmArticle{?}

\setcopyright{rightsretained}

\newcommand{\dX}[2]{\frac {d{#1}}{d{#2}}}
\newcommand{\pdX}[2]{\frac {\partial{#1}}{\partial{#2}}}

\newcommand{\pddX}[2]{\frac {\partial^2{#1}}{\partial{#2}^2}}

\newcommand{\delt}{\Delta_t}

\newcommand{\vecvar}[1]{\mathbf{#1}}
\newcommand{\tnsvar}[1]{\bm{#1}}

\newcommand{\uv}{\vecvar{u}} %
\renewcommand{\vv}{\vecvar{v}} %
\newcommand{\wv}{\vecvar{w}} %
\newcommand{\fv}{\vecvar{f}} %
\newcommand{\gv}{\vecvar{g}} %
\newcommand{\bv}{\vecvar{b}} %
\newcommand{\cv}{\vecvar{c}} %
\newcommand{\xv}{\vecvar{x}} %
\newcommand{\pv}{\vecvar{p}} %
\newcommand{\qv}{\vecvar{q}} %
\newcommand{\yv}{\vecvar{y}} %
\newcommand{\psiv}{\bm{\uppsi}} %
\newcommand{\lambdav}{\bm{\uplambda}} %

\newcommand{\sigt}{\tnsvar{\sigma}} %
\newcommand{\taut}{\tnsvar{\tau}} %
\newcommand{\lambdat}{\tnsvar{\lambda}} %
\newcommand{\omegat}{\tnsvar{\omega}} %

\newcommand{\Ct}{\tnsvar{C}} %
\newcommand{\Ft}{\tnsvar{F}} %
\newcommand{\St}{\tnsvar{S}} %
\newcommand{\Rt}{\tnsvar{R}} %
\newcommand{\Qt}{\tnsvar{Q}} %

\newcommand{\R}{\mathbb{R}}

\DeclareMathOperator{\Sym}{Sym}
\DeclareMathOperator{\Skew}{Skew}
\DeclareMathOperator{\SO}{SO}

\newcommand{\Lts}{L^2_{\Omega}} %

\newcommand{\Vs}{V_{\Omega}} %
\newcommand{\Ts}{T_{\Omega}} %
\newcommand{\SOs}{\SO_{\Omega}} %
\newcommand{\Syms}{\Sym_{\Omega}} %
\newcommand{\Skews}{\Skew_{\Omega}} %

\newcommand{\Id} {\mathbb{I}_3}

\newcommand{\genloss}{\mathcal{L}}
\newcommand{\phyzloss}{\genloss_{\text{phys}}}
\newcommand{\FCloss}{\genloss_{\text{FC}}}
\newcommand{\edgelenloss}{\genloss_{\vert e\vert}}

\newcommand{\typstiff}{\hat{\sigma}}
\newcommand{\typlength}{\hat{L}}
\newcommand{\typacc}{\hat{g}}

\newcommand{\testbasis}{\mathcal{B}_P^d}
\newcommand{\trilinear}{N^\varphi}

\begin{document}

\title{Neurally Integrated Finite Elements for Differentiable Elasticity on Evolving Domains}

\author{Gilles Daviet}
\affiliation{%
  \institution{NVIDIA}
  \country{France}}
\email{gdaviet@nvidia.com}
\author{Tianchang Shen}
\affiliation{%
  \institution{NVIDIA}
  \country{Canada}}
\email{frshen@nvidia.com}
\author{Nicholas Sharp}
\affiliation{%
  \institution{NVIDIA}
  \country{USA}}
\email{nsharp@nvidia.com}
\author{David I. W. Levin}
\affiliation{%
  \institution{NVIDIA}
  \country{Canada}}
\email{dlevin@nvidia.com}

\begin{teaserfigure}
  \vspace*{-1em}
  \includegraphics[width=\textwidth]{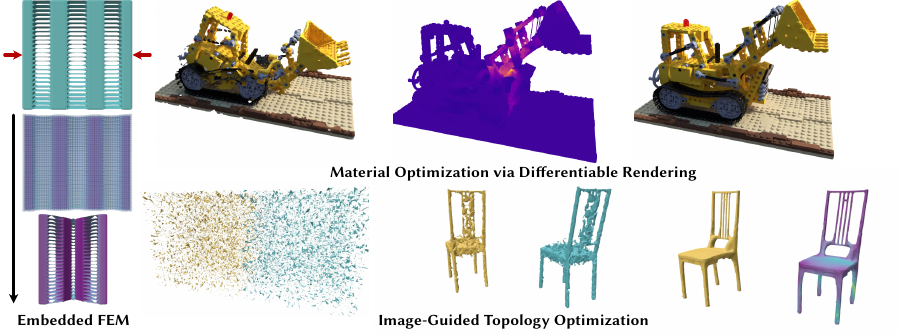}
  \vspace*{-2em}
  \caption{
  Our neurally integrated, high-order mixed finite element solver (left) can accurately simulate evolving implicit geometry (including sub-grid features). It is end-to-end differentiable and can be easily combined with other differentiable tools to enable new applications such as image-guided material, shape and topology optimization (top, bottom).
  }
  \Description{its the teaser}
  \label{fig:teaser}
\end{teaserfigure}

\begin{abstract}
We present an elastic simulator for domains defined as evolving implicit functions, which is efficient, robust, and differentiable with respect to both shape and material. This simulator is motivated by applications in 3D reconstruction: it is increasingly effective to recover geometry from observed images as implicit functions, but physical applications require accurately simulating and optimizing-for the behavior of such shapes under deformation, which has remained challenging. Our key technical innovation is to train a small neural network to fit quadrature points for robust numerical integration on implicit grid cells. When coupled with a Mixed Finite Element formulation, this yields a smooth, fully differentiable simulation model connecting the evolution of the underlying implicit surface to its elastic response. We demonstrate the efficacy of our approach on forward simulation of implicits, direct simulation of 3D shapes during editing, and novel physics-based shape and topology optimizations in conjunction with differentiable rendering.
\end{abstract}



\maketitle

\section{Introduction}
For more than a decade, computer vision has been making strides in improving the output fidelity of reconstruction algorithms. With the advent of differential rendering, it is now possible to create highly complex, three-dimensional geometry from two-dimensional images. The robustness and ease-of-use of these modern methods means that almost any complex real world shape can now be cast as a convincing geometric digital twin.

Increasingly, there is a demand to  use this geometry in application spaces where beyond its shape, its physical responses and robustness are critically important. For instance, engineers may wish to ensure that a captured 3D bracket can support a certain load if fabricated, and roboticists  want to ensure that a captured chair won't collapse under the weight of their robot in a simulated training environment. 

To meet such physical constraints requires not just a geometric reconstruction solution, but one that is physically-aware. While shape optimization and system identification methods have been explored in the past, systems that can optimize over geometry, topology and material properties simultaneously are relatively unexplored. In engineering, most shape optimization methods are strongly model-driven often requiring an initial  parametric shape model (something that existing reconstruction methods do not produce) while system identification approaches assume the input geometry is fixed and optimize only for physical parameters. There is a great need for algorithms that can optimize shape, topology, and material properties holistically, to maximize physical performance or improve response to physical inputs such as forces. 

One approach to this problem is to use a differentiable elasticity simulator as a physical prior in conjunction with a more standard geometry reconstruction method. But state-of-the-art reconstruction approaches generate geometry that is rapidly evolving and can degenerate during the reconstruction process. This, combined with material stiffness parameters that can vary by up-to-four orders of magnitude (at times during optimization) across the object means that robust, in-the-loop simulation for reconstruction is a non-trivial task.

We propose an algorithm that directly attacks these difficulties for elastically deformable objects. Our simulator is built around a regular grid discretization and represents geometry as an implicit function on that same grid. We perform dynamic and quasi-static simulation using a mixed finite-element method (FEM) which supports high-order basis functions if necessary, and prevents performance degradation even when material properties are wildly varying. Crucially, we introduce a neural-network approach to per-element quadrature which allows for smooth, differentiable integration of field quantities across the implicitly-defined domain --- even as it evolves during the reconstruction procedure. 

We combine our novel simulator with the FlexiCubes~\citep{Shen23} reconstruction algorithm and demonstrate its ability to directly produce geometry that is physically reinforced as to avoid excessive deformation under load. We show that the method requires no strong shape prior, divines the geometry and topology of the output as part of the reconstruction process, and can simulate the effect of thin, sub-grid features. Finally, we show how each part of our method (mixed FEM, neural quadrature) is required to achieve stable and robust results.

\section{Related Work}

Geometry reconstruction algorithms focus on producing a consistent 3D representation of a shape from a variety of scanned or synthetic inputs. These inputs include but are not limited to photographs, rendered images, or scan data~\citep{choy20163dr2n2}. These include algorithms for producing triangle meshes~\citep{Liu:Paparazzi:2018, gkioxari2019mesh,liu2024meshformer}, implicit surfaces~\citep{park2019deepsdf, mittal2022autosdf}, point clouds~\citep{fan2017point, wu2020pq}, Gaussian splats~\citep{zhang2024gs,charatan2024pixelsplat} and NeRFs~\citep{mildenhall2020nerf, yu2021pixelnerf, hong2023lrm}.  Our work focuses on providing a compatible, elastodynamics simulator that seamlessly augments geometric reconstruction algorithms to enable the production of physically-sound reconstructed geometry. 

Shape and Topology Optimization~\citep{Bendsoe:2009:Topology,NEURIPS2021_55d99a37,ALLAIRE:2004:SO} are related problems from the engineering literature. While they still optimize for output geometry, they seek to optimize the shape (resp. topology) of an object with respect to some physical properties (e.g, compliance)~\citep{WANG:2003:LS} rather than purely seeking geometric or visual agreement with input. The boundary between these methods is somewhat indistinct since most topology optimization schemes can alter shape, via adding or removing material, hence shape optimization more often relies on a parameterized shape template to constrain results to a design space (e.g, \citet{Panetta2017}).

Finally, systems identification problems endeavour to identify material parameters that match observed motion and/or deformation. Typically an existing, parameterized material model is assumed and differentiable simulation is used to ascertain the parameters that harmonize simulated and observed object behavior~\citep{li2023pac,Huang24}. Other methods avoid differentiable simulation via techniques such as modal analysis~\citep{DAC2017} but in all cases the geometry is known prior to the physics parameter optimization. For instance methods such as PAC-NeRF~\citep{li2023pac} first estimate geometry from images then perform system identification on that fixed geometry. Practically this means that changing geometry cannot be used to optimize physical behavior by construction. In contrast, our novel, neurally-integrated, differentiable elasticity solver is fully differentiable with respect to geometry and material parameters enabling both image-driven and mechanically optimized reconstruction seamlessly and simultaneously.

 Differentiable simulations are well-studied and exist for optimizing the trajectory of rigid objects~\citep{popovic2000interactive}, fluids~\citep{mcnamara2004fluid}, coupled rigid and fluid motion~\citep{10.1145/3618318} and deformable objects~\citep{gradsim,du2021_diffpd,hu2019difftaichi, Geilinger2020add}, but less frequently support shape derivatives. Topology optimization schemes rely on meshless methods~\citep{li2020meshless} or finite elements~\citep{Schumacher:2015:MCE,GAIN2015411} to compute necessary physical responses. Crucially, these previous approaches all suffer from one or more issues that make them less than ideal for general geometric  optimization tasks. Standard, conforming mesh FEM (typically applied on tetrahedral or hexahedral elements) requires high mesh resolutions~\citep{liu2018narrow} to capture the correct behavior of the complex geometries generated by the optimization process. Simulation methods which rely on evolving meshes can require difficult and time-consuming remeshing operations~\citep{wicke2010dynamic, Misztal2012dsc, Huang24} to keep elements well-conditioned, while density-based meshless methods yield a fuzzy interface~\citep{li2020meshless}.

Implicit functions and their neural counterparts have become the \textit{de facto} geometry representation for shape optimization due to their ability to compactly represent complex evolving geometries~\citep{gao2020deftet,Shen23}. We observe that trying to directly and exactly represent these functions using an evolving, high-resolution mesh is the main source of algorithmic complexity as well as memory and computational pressure. Rather, we are influenced by the success of high-order embedded methods in predicting complex deformations of intricate shapes using simple regular grids~\citep{Longva20}. When coupled with appropriately accurate quadrature schemes~\citep{kim2011fast,Patterson12}, excellent accuracy can be obtained. However, these methods need the underlying simulated geometry \textit{a priori}, meaning it is not obvious how to apply them to applications where the geometry may evolve.

Fixed quadrature points and weights introduce non-smoothness and ill-conditioning ~\citep{van2013level} as the surface evolves past them.
Inspired by recent applications of machine-learning techniques in other areas of simulation~\citep{Wang2022musculotendons, Li2023metamaterial, Tymms2020tactile,10.1145/3610548.3618225}, 
our solution is to build a novel finite element method around a neural integration scheme, which uses a small, per-element neural network to learn the quadrature point locations and weights as a function of the underlying implicit shape. This allows points and weights to evolve smoothly, along with the shape itself, yielding higher quality results. Learning functions within grid cells is oft-used in geometry processing (e.g, for mesh extraction~\citep{chen2022neural}) however we believe this is the first time it has been applied to this particular problem. 

Our differentiable simulator is built around a mixed-variational approach to elasticity~\cite{reissner1985mixed, Simo90mfem}. In particular we modify the rotation-aware extension proposed by~\cite{Trusty22} which allows for simulation performance independent of material stiffness --- an important property for system identification where material parameters can range over several orders-of-magnitude. We use a novel, performant variant of the scheme that avoids a per-element singular-value-decomposition at each simulation sub-step and supports high-order elements.

Our major contribution is a differentiable simulator that enables simultaneous optimization of object shape, topology and material properties via drop-in combination with existing geometry reconstruction algorithms. In service of this we develop a novel neural quadrature scheme suitable for finite element algorithms on evolving surfaces, a fast mixed finite-element method to enable robust simulation across wide ranges of material parameters and a gradient preconditioner to improve convergence. 

\section{Neurally integrated FEM}
\label{sec:diff_fem}
We first recall a few preliminaries about FEM and numerical integration.
We consider a material domain $\Omega \subset \R^3$ equipped with a displacement field $\uv$ defined over a space\footnote{typically a subspace of the Sobolev space $H^1(\Omega)^3$} $\Vs$, and write $\Ft(\uv) := \Id + \nabla \uv$ the deformation gradient.

\subsection{Weak-form elasticity}
\label{sec:classic_fem}
The kinetic and potential energies of the system are defined as
\begin{align*}
E_k &:= \int_{\Omega}{\rho \dot{\uv}^2}, & E_p &:= \psi(\Ft(\uv)) - \int_{\Omega}{\uv.\gv}, & \psi(\Ft) &:= \int_{\Omega}{\Psi(\Ft)},
\end{align*}
with $\Psi : \R^{3\times3} \to \R$ the local elasticity potential, and $\gv$ an external force density. 
Using an implicit Euler integrator with timestep $\delt$ (possibly infinite in the quasistatic limit), such that $\dot{\uv} \sim \frac {\uv - \uv^t} {\delt} $, with $\uv^t$ the begin-of-step velocity, the conservation of momentum over the timestep can be expressed as the minimization of the incremental potential~\citep{Kane2000}
\begin{equation}
\label{eq:ip}
\begin{aligned}
    &\min_{\uv \in \Vs} E_t(\uv), 
    & E_t(\uv) &:= \psi(\Ft(\uv)) + \frac 1 2 a(\uv, \uv) - b(\uv), \\
    a(\uv, \vv) &:= \int_{\Omega} \frac{\rho}{\delt^2} \uv . \vv, &
    b(\vv) &:= \int_{\Omega} \left(\gv + \frac \rho {\delt^2} \uv^n \right).\vv.  
\end{aligned}
\end{equation}
Writing the optimality condition $\partial_{\uv} E_t = 0$ as directional (Gâteaux) derivatives over all of $\Vs$ yields the weak-form FEM formulation,
\begin{equation}
\label{eq:weak_cfem}
 a(\uv, \vv) + \int_{\Omega}\pdX{\Psi}{\Ft} \left( \Ft(\uv) \right) : \nabla \vv = b(\vv) \quad \forall  \vv \in \Vs.
\end{equation}

We can then choose a finite subspace for $\Vs$, typically polynomials defined over the elements of a mesh $\mathcal{M}$, and write Equation~(\ref{eq:weak_cfem}) for each function $\vv_i$ of our discrete basis. This yields as many nonlinear scalar equations, that can be solved for instance using a quasi-Newton method~\citep[e.g,][]{Smith18}.
Doing so assumes being able to evaluate integrals over $\Omega$, or in practice over any element $K$ of the mesh $\mathcal{M}$. 
As analytical expressions may not be available, we resort to approximate quadrature rules, that is, sets of points and weights $(w^K_p)$, $(\yv^K_p)$ such that for any polynomial $P$ of degree less than or equal to $d$, $\int_K P = \sum_p w^K_p P(\yv^K_p)$. The integer $d$ is called the \emph{order} of the formula. Quadrature rules have been derived and tabulated for common elements (such as tetrahedra and hexahedra) at all practical polynomial orders~\citep{Cools2003}.
However, standard quadrature rules are only applicable when the mesh $\mathcal{M}$ coincides with the material domain $\Omega$, and generating good-quality conforming volumetric meshes from a surface is both expensive and not easily differentiable. In the next paragraphs we show how we can avoid building such a mesh altogether and cheaply generate good non-conforming quadrature formulas for surfaces that are implicitly defined by a signed-distance function (SDF) discretized over a hexahedral mesh --- such as in marching cubes~\cite{lorensen1998marching}, OpenVDB~\cite{museth2013openvdb}, or FlexiCubes~\cite{Shen23} grids.

\begin{figure}
    \centering
    \includegraphics[width=.85\linewidth]{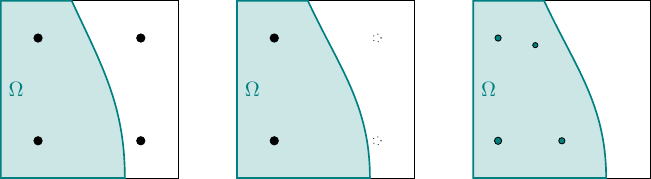}
    \caption{Illustration in 2D of order-2 quadrature points for a boundary voxel for, from left to right, Full, Clip and Neural quadrature formulae. The neural quadrature points and weights are updated smoothly when the boundary moves, while other formulas experience jumps.}
    \label{fig:quadratures}
\end{figure}

\subsection{Optimized Quadrature Rules}
\label{sec:optim_qp}
A first observation is that for the weak-form FEM described in Equation~({\ref{eq:weak_cfem}}), we do not need the exact domain geometry; we only need the capacity to numerically integrate functions over elements with good accuracy. 
Let us consider a mesh element $K$ and a domain $\Omega$ such that $K \nsubseteq \Omega$; we want to compute integrals over the part of $K$ where there is material, i.e. $K \cap \Omega$. One first possibility, which we will refer to as the Clip quadrature -- see Figure~\ref{fig:quadratures}, would be to multiply the integrand, or equivalently the quadrature point weights, with the domain indicator function $\chi_{\Omega}$,
\begin{align*}
\int_{K \cap \Omega} f &= \int_K f \chi_{\Omega} \sim \sum_p w_p \chi_{\Omega}(\yv_p) f(\yv_p),&  \chi_{\Omega} &:= \left\{ 
\begin{aligned}
1 & \text{ on } \Omega, \\
0 & \text{ elsewhere.}
\end{aligned}\right.
\end{align*}

However, the indicator function $\chi_\Omega$ is highly nonlinear and the quadrature quality will quickly degrade, leading to unstable simulations. Moreover, $\chi_\Omega$ is non-differentiable with zero gradient almost everywhere, hindering the computation of meaningful derivatives of the integration result with respect to the domain. Instead, following~\citet[e.g,][]{Patterson12}, we opt to derive new quadrature points and weights that can accurately integrate polynomials at a chosen order $d$ on the actual material domain. 
Using an optimization point of view, we express this quest as 
\begin{align}
\label{eq:min_qp}
\min_{w_p, \yv_p} \mathcal{Q}_K,\quad \mathcal{Q}_K := 
\sqrt{ \sum_{P \in \testbasis} \left( \int_{K \cap \Omega} P - \sum_p w_p P(\yv_p)  \right)^2},
\end{align}
with $\testbasis$ a basis for polynomials of degree $d$. 
Monomials may be used to define $\testbasis$, in which case minimization~(\ref{eq:min_qp}) is known as moment fitting~\citep{Bremer10momentfitting}. For symmetry reasons, we prefer to use Lagrange polynomials defined on the Lobatto--Gauss--Legendre nodes as our basis.

\begin{figure}
    \centering
    \includegraphics[width=.3\linewidth]{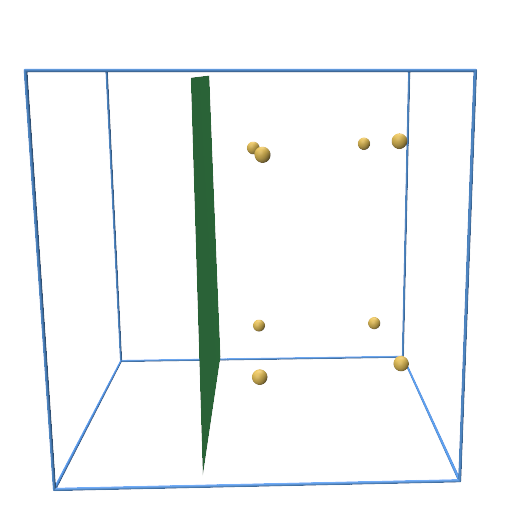}
    \includegraphics[width=.3\linewidth]{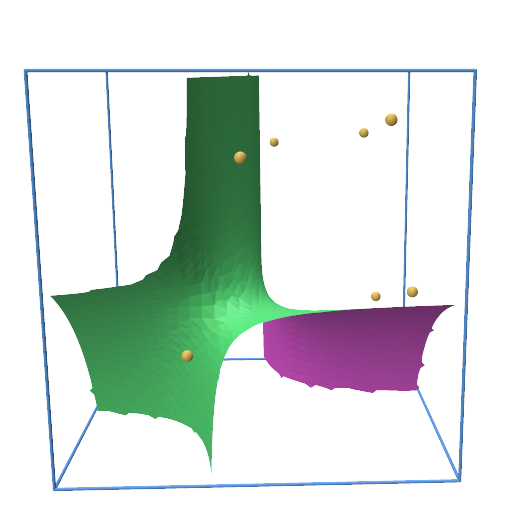}
    \includegraphics[width=.3\linewidth]{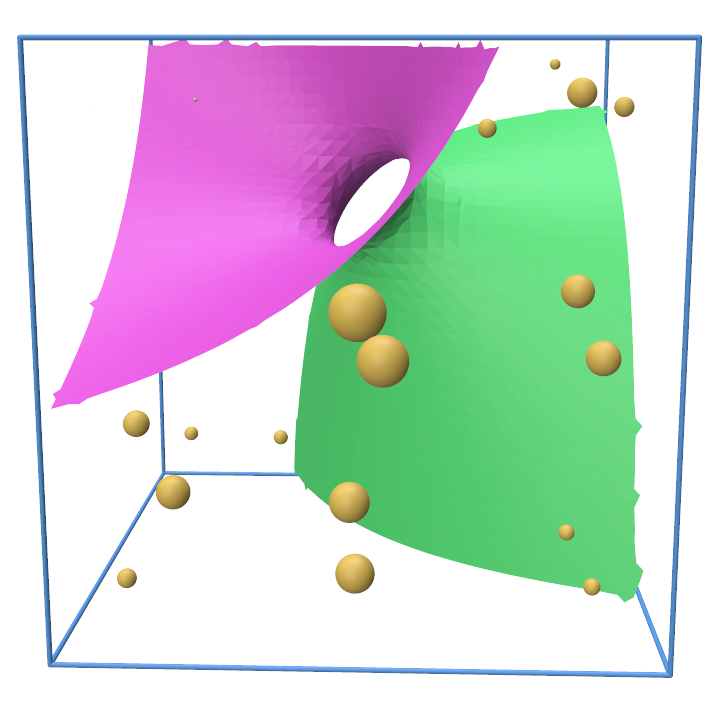}
    \caption{Learned quadrature points (yellow) for integrating over the part of the unit voxel defined by trilinear interpolation of the corner SDF values (visualized by the green isosurface). Left and middle depict 8-point order-2 quadrature, right is 27 point order-4. Size is proportional to the quadrature point weight.}
    \label{fig:qp}
\end{figure}

\subsection{Neural Quadrature Rule Prediction}
\label{sec:neural_quad} 
In our settings of interest, the material domain $\Omega \cap K$ is implicitly defined 
as the region of $K$ where the SDF $\varphi^K$ is negative. Moreover, the function $\varphi^K$ is itself discretized as a finite set of nodal values $(\varphi^K_j)$,
$
\varphi^K(\xv) = \sum_j \varphi^K_j \trilinear_j(\xv),
$
with the shape functions $(\trilinear_j)$ assumed identical for all nodes --- trilinear in our case\footnote{As we are only concerned with the 0-isosurface, the discretization does not need to preserve the eikonal property of the SDF.}. 
Our problem thus reduces to finding, from a set of input SDF values $(\varphi^K_j)$, quadrature points $(\yv^K_p)$ and weights $(w^K_p)$ that are  an approximate solution to the minimization problem~(\ref{eq:min_qp}).

In principle, one could use direct numerical optimization techniques. 
However, we want the following desirable properties for our quadrature rule generation scheme: 
\begin{enumerate}[label=(\alph*)]
    \item it should be extremely cheap, as it will need to be performed for every partially filled element of the mesh, each time the boundary is evolved within a shape optimization loop;
    \item the resulting $(\yv^K_p, w^K_p)$ should be continuous with respect to $\varphi^K_j$, with easily-accessible and well-behaved gradients; 
    \item the number of quadrature points should be fixed, both for controlling the cost of the simulation and, in Mixed FEM settings, for satisfying an element-compatibility condition;
    \item for numerical conditioning the ratio of weights between the different quadrature points should be limited.
\end{enumerate}
On the other hand, our applications do not require
\begin{enumerate}[label=(\alph*)]
\setcounter{enumi}{4}
\item the quadrature rule to be extremely accurate,
\end{enumerate}
as our object reconstruction objective implies uncertainty about the exact location of the domain boundary anyway.

Most moment-fitting approaches~\citep{Patterson12, Bremer10momentfitting, Longva20} aim to minimize the nonlinear problem~(\ref{eq:min_qp}) to high accuracy, i.e, achieve~(e) at the detriment of~(a) and often~(c). Moreover~(b) is usually out of reach for non-convex optimization problems with local minima.
\citet{Muller13} keep the number and position of the quadrature points fixed and optimize for the weights only, yielding a linear problem that achieves both~(b) and~(c). However, the resulting weights may be null or negative, contradicting~(d), and this restricted optimization space limits accuracy for a given number of quadrature points --- see also Appendix~\ref{app:training}.

Instead, we propose to train a small neural network to learn the mapping $(\varphi_j) \mapsto (\yv_p, w_p)$. 
Precisely, we fit a network which takes as input a stacked vector of implicit SDF values at a single cell's corners, and outputs the quadrature point locations and weights 
within the cell. 
At simulation time we simply need to run inference for all current voxels, stacked as a single tensor; this is extremely cheap, achieving~(a). Criteria~(b) and~(c) are satisfied by construction, and conditioning~(d) can be controlled as an additional training loss term.
We train this network once for integration order $2$ and $4$, and use it for all experiments.

\paragraph{Architecture, Losses, and Training}
We choose the network architecture to be a simple multilayer perceptron with $N_{MLP}=5$ fully-connected layers of size $W_{MLP}=64$ for order 2 and $W_{MLP}=128$ for order 4, and ReLU activations on hidden layers.
Network inputs are normalized such that the gradient of $\varphi$ is unit at the cell center, and network outputs are parameterized as offsets from and multipliers for the usual Gauss--Legendre points and weights. 
We define the loss function as the sum 
\begin{equation}
\mathcal{L}_\textrm{QuadNet} := 
\mathcal{Q}_K + 
10^{1} \mathcal{Q}_{\square} + 
\gamma_{\star} \mathcal{Q}_{\star}
\end{equation}
where $\mathcal{Q}_K$ is from Equation~(\ref{eq:min_qp}), with target integrals $\int_{\Omega \cap K} \testbasis$ computed using brute force uniform integration at high resolution, $\mathcal{Q}_{\square}$ is quadratic barrier enforcing  quadrature coordinates to stay in $[0, 1]^3$, and $\mathcal{Q}_{\star}$ is a conditioning term penalizing the log ratio of the maximum to minimum quadrature weight.
We generate a training set of $2^{24}$ voxels, each consisting of $8$ random corner SDF values, and train for 64k iterations with the AdamW optimizer with batch size $2^{18}$ chosen as to fit our specific GPU memory. Training the order-2 network takes about $1.5$ hours on a NVIDIA GeForce RTX 3080Ti GPU --- which was also used to generate all of the examples in the remainder of this article --- while training the order-4 network takes about $30$ hours on a NVIDIA A40 GPU. Figure~\ref{fig:qp} shows the inferred quadrature points for selected voxel configurations, and  Appendix~\ref{app:training} provides more evaluations and training details.

\section{Neurally Integrated Mixed FEM}
\label{sec:mfem}
Mixed FEM consists in discretizing the elasticity equations over multiple fields, rather than just the displacement $\uv$, which can significantly improve numerical convergence properties~\citep{BrezziFortin91, Simo90mfem, Ko2017mfem, Francu2021locking}. This is of particular interest to us as we want to embed our solver in a shape reconstruction loop, and as such, desire to obtain a good approximation of the final result even when truncating the solve to a few Newton iterations and regardless of the material stiffness.

The rotation-aware Mixed FEM formulation described by \citet{Trusty22} boasts this property, however as presented is limited to linear displacements and piecewise-constant strains and stresses. Below we propose a four-field extension of this mixed formulation to arbitrary finite elements, and show how it can be used in conjunction with our Neural Quadrature integration strategy. Unless otherwise mentioned, this formulation will serve as the basis for our differentiable elasticity simulations.

\subsection{Generalized four-field Mixed FEM}
We denote by $\Ts$ the space of square-integrable $3\times3$ tensor fields, and define $\SOs$, $\Syms$ and $\Skews$, the subspaces of $\Ts$ whose values are rotations, symmetric tensors, and skew-symmetric tensors, respectively.
We introduce two additional primal fields, the symmetric strain $\St \in \Syms$ and rotation $\Rt \in \SOs$, related to the deformation gradient through the constraint $\Ct(\uv, \Rt, \St) := \Ft(\uv) - \Rt \St = 0$. Being rotation-independent, the local elastic potential $\Psi$ can now be measured directly on $\St$ rather than $\Ft$.
Minimization of the incremental potential~(\ref{eq:ip}) can be expressed as the constrained optimization
$$
\min_{\begin{array}{c}\uv \in \Vs, \St \in \Syms, \Rt \in \SOs\\\Ct(\uv, \Rt, \St)=0\end{array}}  \frac 1 2 a(\uv, \uv) - b(\uv) + \psi(\St)
$$
or equivalently as a saddle point of the associated Lagrangian,
\begin{equation*}
\label{eq:lagrangian}
\begin{aligned}
&\min_{\uv \in \Vs, \St \in \Syms, \Rt \in \SOs} \max_{\sigt \in \Ts} \mathcal{L}(\uv, \St, \Rt, \sigt), \\
\mathcal{L}(\uv, \St, \Rt, \sigt) &:= \frac 1 2 a(\uv, \uv) - b(\uv) + \Psi(\St) + c(\uv, \St, \Rt, \sigt), \\
 c(\uv, \St, \Rt, \sigt) &:= \int_{\Omega} \Ct(\uv, \Rt, \St) : \sigt.
\end{aligned}
\end{equation*}
Solutions of problem~(\ref{eq:ip}) must thus satisfy $\partial \mathcal{L} = 0$, that is,
\begin{align}
    \label{eq:nl_cm0}
    a(\uv, \vv) + c_{,\uv}(\vv, \sigt) - b(\vv) &= 0 & \forall \vv &\in \Vs, \\
    \label{eq:nl_elast0}
    \psi_{,\St}(\St; \taut) + c_{,\St}(\Rt; \taut, \sigt) &= 0 & \forall \taut &\in \Syms, \\
    \label{eq:zero_skew0}
    c_{,\Rt}(\St; \Qt, \sigt) & = 0 & \forall \Qt &\in \SOs, \\
    \label{eq:nl_cst0}
    c(\uv, \Rt, \St, \lambdat) &= 0 & \forall \lambdat &\in \Ts,
\end{align}
where the forms $\psi_{,\St}$ and $c_{,q}$ are directional derivatives, i.e,
$$
\begin{aligned}
\psi_{,\St}(\St; \taut) &:= \int_{\Omega} \pdX{\Psi}{\St}(\St) : \taut, &
c_{,\uv}(\uv, \lambdat) &:= \int_{\Omega} \nabla \uv : \lambdat, \\
c_{,\Rt}(\St; \Rt, \lambdat) &:= \int_{\Omega} \Rt \St : \lambdat, &
c_{,\St}(\Rt; \St, \lambdat) &:= \int_{\Omega} \Rt \St : \lambdat. \\
\end{aligned}
$$
Note that at equilibrium the Lagrange multiplier $\sigt$ coincides with the first Piolat--Kirchhoff stress tensor, $\pdX{\Psi}{\Ft}$~\citep{BonetWood08}.

Unfortunately, directly applying Newton iterations to Equations~(\ref{eq:nl_cm0}--\ref{eq:nl_cst0}) would lead to numerical difficulties. Indeed, the elasticity Hessian may be indefinite, and there is no coercive potential for the rotation variable; see Appendix~\ref{app:mfem} for details. To remedy to this problem, we re-inject the constraint $\Ct$ into Equations~(\ref{eq:nl_elast0}--\ref{eq:zero_skew0}) using an Augmented--Lagrangian-like penalization term $\epsilon$, 
\begin{align}
    \label{eq:nl_cm}
    a(\uv, \vv) + c_{,\uv}(\vv, \sigt) - b(\vv) &= 0 & \forall \vv &\in \Vs \\
    \label{eq:nl_elast}
    \psi_{,\St}(\St; \taut) + c_{,\St}\left(\Rt; \taut, \sigt + \varepsilon C\left(\uv, \Rt, \St\right) \right) &= 0 & \forall \taut &\in \Syms \\
    \label{eq:zero_skew}
    c_{,\Rt}\left(\St; \Qt, \sigt+ \varepsilon C\left(\uv, \Rt, \St\right)\right) & = 0 & \forall \Qt &\in \SOs \\
    \label{eq:nl_cst}
    c(\uv, \Rt, \St, \lambdat) &= 0 & \forall \lambdat &\in \Ts.
\end{align}
The penalization parameter $\varepsilon$ has the dimension of an elastic modulus, and in practice we set it equal to the typical stress $\typstiff := \rho \typacc \typlength$, with $\typlength$ and $\typacc$ typical length and acceleration, respectively.
We proceed to solve system~(\ref{eq:nl_cm}--\ref{eq:nl_cst}) using projected Newton iterations; we describe how to do so efficiently in Appendix~\ref{app:mfem}. Differences with the original approach from~\citet{Trusty22} are outlined in section~\ref{sec:mfem_perf}.

\subsection{Combination with Neural Quadrature}
Our Mixed FEM solver does not overly restrict the choice of quadrature formulas, as long as they are of sufficient accuracy. As outlined in Appendix~\ref{app:discrete_mfem}, it mandates for efficiency that the quadrature points used to integrate Equations~(\ref{eq:nl_elast}--\ref{eq:nl_cst}) coincide with the degrees of freedom of the strain spaces; but we can freely  pick the location of those Lagrange polynomial nodes. 
We can thus combine the Mixed FEM formulation with our Neural Quadrature from Section~\ref{sec:neural_quad}. For hexahedral elements we use polynomials of similar degree $k$ for the displacement and tensor spaces, meaning that we can use the same formula of order $d=2k$ for all of our integrals. While the displacement space $\Vs$ is continuous with nodes positioned according to the mesh, the strain, rotation and stress spaces use a discontinuous Lagrange polynomial basis with nodes collocated to the quadrature points $(\yv_p^K)$ inferred in each element. Note that in practice the tensor fields will only be evaluated at said nodes, so we do not need to consider general interpolation.

\section{Forward simulation results}

We have implemented our FEM and Mixed FEM solvers using the \texttt{warp.fem} module from the NVIDIA Warp~\citep{warp2022} library, which allows us to conveniently express the linear and bilinear forms described in sections~\ref{sec:classic_fem} and~\ref{sec:mfem_continuous} and provides auto-differentiated numerical integration code with respect to all of the domain and material parameters.
Below we assess the efficiency of neural quadrature and Mixed FEM solver, first on a simple dumbbell geometry then on more complex topologies.

\subsection{Dumbbell}
\label{sec:dumbbell}
\begin{figure}
     \centering
     \includegraphics[width=.485\linewidth]{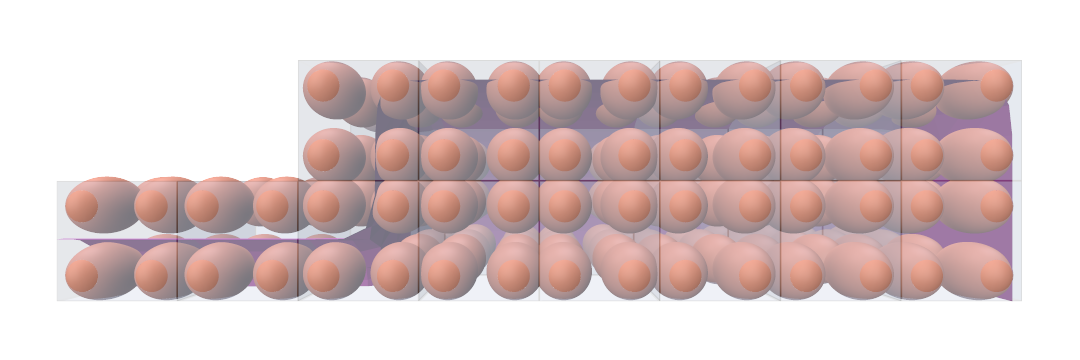}
     \includegraphics[width=.485\linewidth]{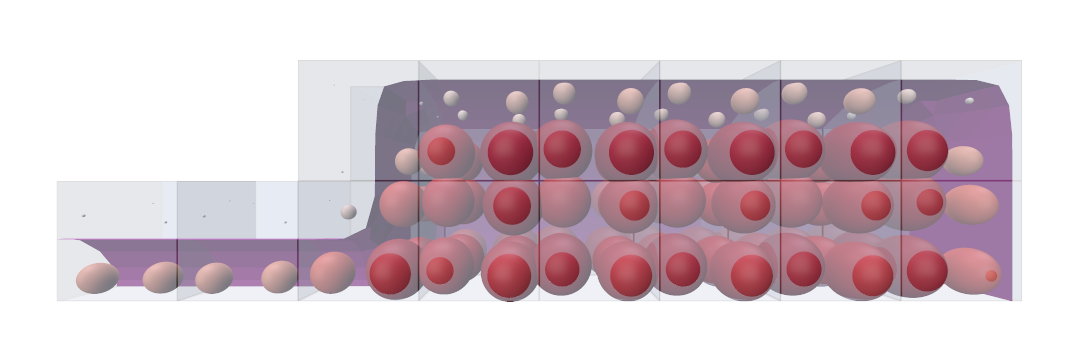}
     \includegraphics[width=.485\linewidth]{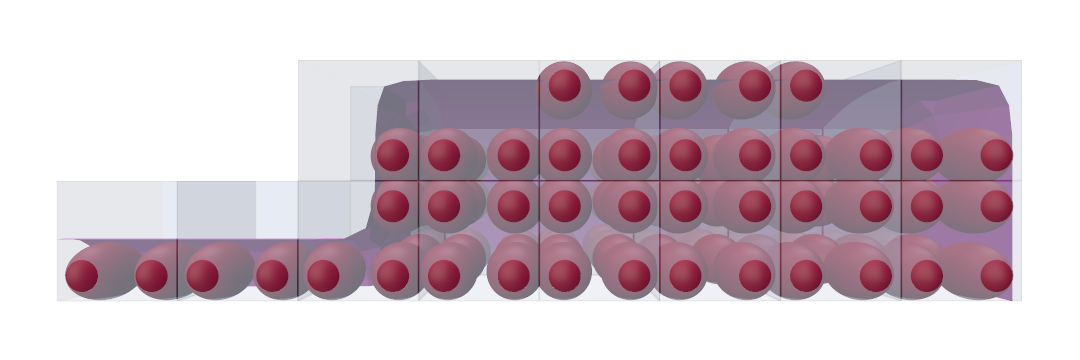}
     \includegraphics[width=.485\linewidth]{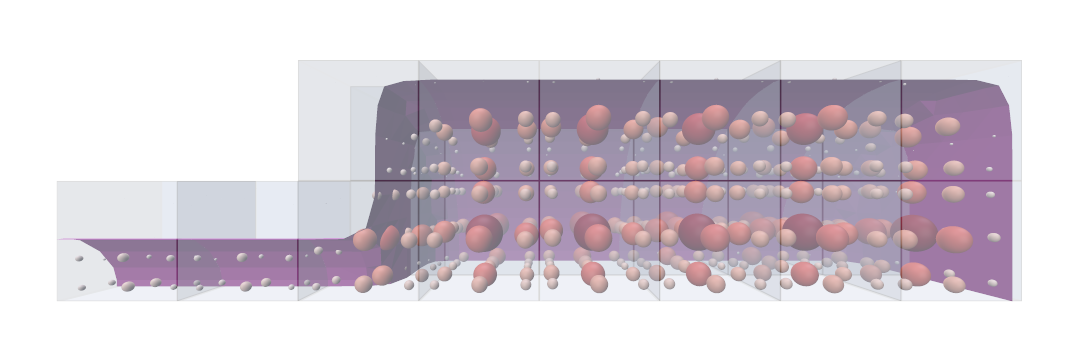}
      \caption{
      Visualization of quadrature points generated for a SDF discretized on a grid at resolution $16^3$, using order-$2$ Full (top left) and Clip (bottom left) quadratures, and our Neural quadrature at order $2$ (top right) and $4$ (bottom right). The non-empty voxels are shown in blue, and the SDF iso-surface (extracted at resolution $64^3$) is shown in purple. For clarity, only one octant is shown.
      }
     \label{fig:qp_embed}
 \end{figure}

We define a continuous dumbbell SDF as the union of three analytical cylinders, the middle one being of smaller radius than the other two. We then discretize this SDF on regular grids at resolutions varying from $8^3$ to $64^3$. 
We generate quadrature formulas of order 2 (8 points) and 4 (27 points) from this discrete SDF using the Full (regular Gauss--Legendre points and weights), Clip (filtering-out points in the SDF exterior), and our Neural approaches to perform both displacement-only and Mixed FEM simulations. The deformation is visualized by embedding the isosurface extracted using dual marching-cubes at $64^3$ resolution within the simulation grid, as illustrated in Figure~\ref{fig:qp_embed}. We use the Stable Neo--Hookean elastic model from \citet{Smith18} with Poisson ratio $\nu=0.4$.

\begin{figure}
     \centering
     \includegraphics{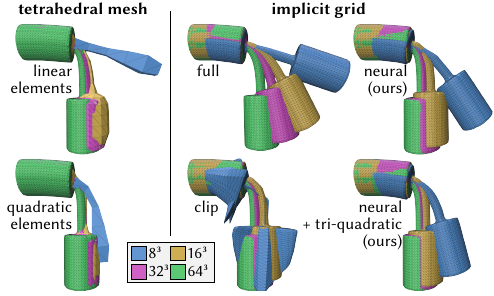}
      \caption{
      Comparison of equilibrium behavior for the dumbbell across several SDF grid resolutions and discretizations.
      \emph{Left}, a tetrahedral mesh is first extracted from the grid via the FlexiCubes algorithm, and simulated with linear and quadratic elements.
      \emph{Right}, our grid-based simulations are performed with full and clipped quadrature on tri-linear elements, as well as our neural quadrature on both tri-linear and tri-quadratic elements. 
      Color denotes the grid resolution at which the mesh is extracted or the simulation is performed, respectively.
      The grid simulation is interpolated to a high-resolution surface for visualization.
      }
     \label{fig:qp_comp}
 \end{figure}

\paragraph{Cantilever} 
The first experiment consists in clamping one end and letting the (soft) dumbbell sag under gravity, studying the impact of the grid resolution on the achieved equilibrium shape for different quadrature formulae. We also compare our non-conforming approach with the simulation of a tetrahedral mesh generated from the discrete grid using the algorithm from \citet{Shen23}.
Figure~\ref{fig:qp_comp} shows the equilibrium shapes obtained using our Mixed FEM formulation --- we also performed this experiment with displacement-only FEM and obtained visually identical results. At the fine $64^3$ resolution, all experiments converge to the same shape. At coarse resolutions however, the Full quadrature and the linear tetrahedral mesh underestimate the deformation the most, while the Clip quadrature suffers from instabilities. The quadratic tetrahedral mesh and the tri-quadratic embedded simulation (using order-4 neural quadrature) perform best, followed by the order-2 neural quadrature with trilinear displacements.

\begin{figure}
     \centering 
     \includegraphics{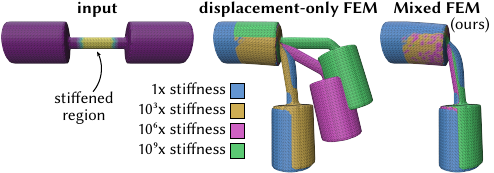}
      \caption{
      At high stiffness ratio, displacement-only FEM suffers from slow convergence, while Mixed FEM does not.
      Here, the center region of the dumbbell is stiffened by an increasing factor, and in each case the Newton loop is truncated after 250 iterations.
      }
     \label{fig:mfem_comp}
 \end{figure}
 
\paragraph{Large stiffness ratios} 
We now increase the elastic modulus of the middle region of the dumbbell (Figure~\ref{fig:mfem_comp}) with stiffness ratios up to $10^9\times$. At lower ratios, displacement-only and Mixed FEM perform identically, but for the higher ratios classic FEM suffers from high damping of the rotational mode in the stiff regions, and remains far from the converged shape even after a $250$ Newton iterations.
This is consistent with the results from \citet{Trusty22}.
 
\begin{figure}
     \centering
     \includegraphics{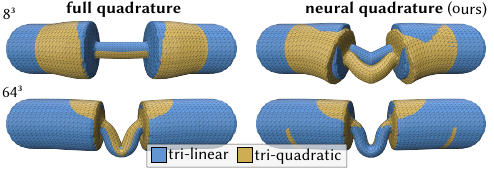}
      \caption{Comparison of buckling shapes across resolution (\emph{by row}), element types (\emph{by color}), and quadrature strategies (\emph{by column}).
      }
     \label{fig:degree_comp}
 \end{figure}
 
\paragraph{Buckling} 
Finally, Figure~\ref{fig:degree_comp} looks at the impact of changing the polynomial degree of the displacement field for the Full and Neural quadrature formulae, at $8^3$ and $64^3$ resolution. 
We observe little impact for the coarse Full and fine Neural buckling simulations; the former fails to take into account the thinner part for both degrees, with the latter results in identical converged shapes. The tri-quadratric displacement is most interesting for the coarse Neural simulation, with an equilibrium shape much closer to the high-resolution solution than with tri-linear displacements.

\subsection{Complex geometries}

\begin{figure}[h]
     \centering
     \includegraphics[width=\linewidth]{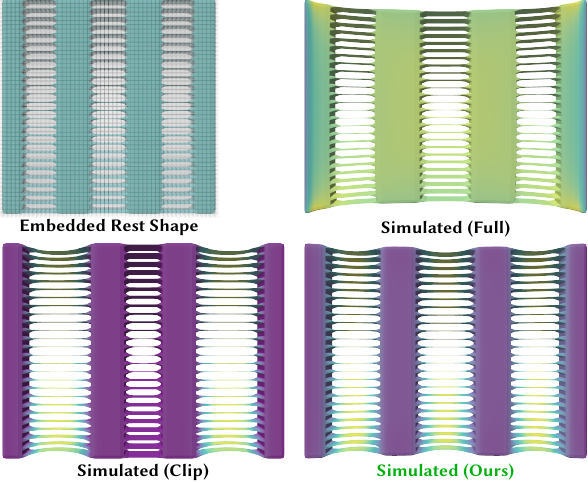}
     \caption{A slab with sub voxel-sized features simulated with Full, Clip and Neural  quadrature formulas. Shading denotes the norm of the strain tensor $S$ relative to the current configuration.}
     \label{fig:aniso}
 \end{figure}

\paragraph{Heterogeneous material}
We simulate a slab of material with heterogeneities roughly the size of one voxel, so that the embedding grid is effectively dense (top left). As shown in Figure~\ref{fig:aniso}, using the regular Full quadrature, the material behaves as if it was homogeneous, with globally uniform strain. The Clip quadrature also yields incorrect behavior, as the strain is no longer transmitted away from the dense clamped regions. Our neural quadrature successfully captures the intricacies of the material at no additional cost.

\begin{figure}
     \centering
     \includegraphics[width=0.85\linewidth]{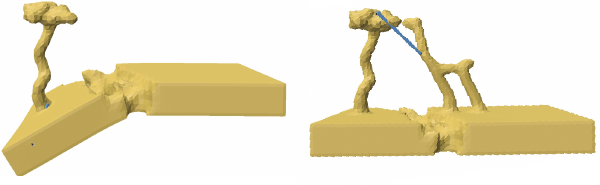}
     \caption{
     Interactive sculpting and simulation of physics-enabled clay
     }
     \label{fig:clay}
 \end{figure}

\paragraph{Interactive editing}
As our framework allows simulation of arbitrarily complex and evolving material topology 
without the need for expensive remeshing, a natural application is a physics-ready virtual playground where the user may interactively add or subtract material and immediately see how it responds to applied forces~(Figure~\ref{fig:clay} and video). 

\section{Physics-aware reconstruction}
\label{sec:phys_aware_recon}

Not only can our neural quadrature handle evolving material domains, it does so in a differentiable way. We exploit this ability to demonstrate physics-aware mesh reconstruction from multiple views. But first we describe how we can efficiently compute the adjoint of our simulations.

\subsection{Simulation Adjoint}
\label{sec:sim_adjoint}
We consider a loss function $\genloss(\pv, \qv)$ to be minimized, with $\pv$ the vector of material and/or shape parameters that we want to optimize, and with $\qv$ the simulation state; $\qv := \uv$ for displacement-only FEM, and $\qv := (\uv, \St, \Rt, \sigt)$ for our Mixed FEM formulation from Section~\ref{sec:mfem}. 
As $\qv$ is the result of a forward simulation, it depends in turn on the parameters $\pv$. Performing gradient-based optimization therefore requires evaluating
$$
\dX{\genloss}{\pv} = \pdX{\genloss}{\pv} + \pdX{\genloss}{\qv} \pdX{\qv}{\pv}. 
$$

One may choose to use full auto-differentiation of the simulator code for all of the above terms. However, evaluating $\pdX{\qv}{\pv}$ requires backtracking through the whole simulation loop --- potentially comprising many solver iterations --- which is costly both in wall time and memory usage. We avoid this overhead by combining auto-differentiated and analytical adjoint computations: by definition, $\qv$ must satisfy an equilibrium condition, either Equation~(\ref{eq:weak_cfem}) for displacement-only FEM or Equations~(\ref{eq:nl_cm}--\ref{eq:nl_cst}) for Mixed FEM. For brevity of notation, let us write this equilibrium condition as $\fv(\pv, \qv) = 0$; the implicit function theorem allows us to express the loss gradient as 
$$
\dX{\genloss}{\pv} = \pdX{\genloss}{\pv} + \underbrace{\pdX{\genloss}{\qv} \left( \pdX{\fv}{\qv} \right)^{-1}}_{\pdX{\genloss}{\fv}} \pdX{\fv}{\pv}. 
$$
As $\pdX{\fv}{\qv}$ can be recognized as the Hessian of the incremental energy potential, computing $\pdX{\genloss}{\fv}$ amounts to solving one linear system similar similar to one Newton iteration from the forward pass\footnote{To compute the exact gradient, we should not perform SPD projection of the elasticity Hessian. However in practice, this allows for a simpler and more efficient solve with little impact on the descent direction, so we use it in the backwards pass as well.}.
The right-multiplication of $\pdX{\genloss}{\fv}$ by $\pdX{\fv}{\pv}$ is then achieved through auto-differentiation of the linear form assembly code, which is directly provided by the \texttt{warp.fem} library~\cite{warp2022} with which our solver is implemented.

Note that in our framework, we do not have to give special treatment to shape derivatives versus material parameter derivatives. The simulator is aware of the material domain through quadrature points and weights, for which the adjoint computation does not require particular considerations. We can then get the derivatives with respect to the implicit surface by backpropagating through the MLP network from Section~\ref{sec:neural_quad}.

\subsection{Physics-aware reconstruction framework}
\label{sec:losses}
We leverage the FlexiCubes~\citep{Shen23} discrete implicit surface representation, which 
consists of SDF values and displacements at nodes of a regular grid, plus per-cell parameters adjusting the isosurface --- effectively, per-cell, per-vertex SDF values with variable vertex positions. This representation has been shown to perform well in conjunction with differentiable rasterization~\citep{Laine2020diffrast}, with stronger ability at capturing sharp features than tet-based alternatives~\citep{shen2021dmtet}. 

Previously ~\citet{Shen23} showed decoupled shape and material optimization, first recovering geometry via FlexiCubes and next optimizing for material properties using a differentiable simulator. In stark contrast we now describe a fully-coupled single stage pipeline wherein gradients from our simulator directly effect reconstructed geometry.
We adjust the physical behavior of the reconstructed object by varying not only material parameters, but also its rest shape; all the while ensuring that renderings of said rest shape remain close to the target.

\paragraph{Loss functions and preconditioning}
On top of the geometric and rendering-based reconstruction losses described by~\cite{Shen23}, which we regroup concisely as $\FCloss$, we add a new physics loss function $\phyzloss$ based on the displacement $\uv$ and stress $\sigt$ fields resulting from the simulation over a timestep $\delt$,
$$
\phyzloss(\delt, \ell_{\uv}, \ell_{\sigt}) := \sqrt[p]{ \int_{\Omega} \frac{1}{\vert \text{det d}\Omega \vert} \left( \ell_{\uv} \left\Vert \uv \right\Vert^p +\ell_{\sigt} \left\Vert \sigt \right\Vert^p \right) },
$$
where $\ell_{\uv}$ and $\ell_{\sigt}$ are constant scaling factors for the displacement and stress terms, and the loss power $p$ allows us to skew the global loss towards either the average or the maximum local loss (in practice we always use $p=8$).
The $\frac{1}{\vert \text{det d}\Omega \vert}$ term scales the local loss inversely to the infinitesimal domain measure to prevent the empty domain from being a trivial optimum.

We emphasize that the $\FCloss$ and $\phyzloss$ losses both affect the shape of the reconstructed model and will oppose each other; tuning the $\ell_{\uv}$ and $\ell_{\sigt}$ coefficient allow biasing the result towards better reconstruction fidelity or physical performance.
Moreover, activating $\phyzloss$ right from the beginning of the optimization is not productive; the initial guess of the FlexiCubes reconstruction consists in random SDF values, leading to many disconnected material pieces, so that running the physical simulation at an such early stage is not meaningful. Instead, we run the first $30\%$ of the optimizer iterations with $\FCloss$ only, then add $\phyzloss$. To ensure a smooth transition, we also increase the simulation timestep $\delt$ progressively.

In practice, performing gradient descent on $\phyzloss$ tends to produces bumpy or fractured surfaces that, while yielding low values of the physics loss, are not visually pleasing. We overcome this issue by preconditioning the grid parameters that are being optimized for (vertex displacement and SDF value) with a smoothing function. To this effect, we apply a convolution with a Gaussian blur kernel before passing those parameters to the FlexiCubes reconstruction and Mixed FEM simulation. 
In a similar fashion, we can optionally enforce symmetry of the optimized shape by applying a symmetric preconditioner to the raw grid parameters.

Finally, adding a loss term $\edgelenloss$ penalizing the total sum of edge lengths of the extracted triangular mesh is helpful for reducing the appearance of unwanted geometry like floaters or protruding details --- under the condition that this term remains small compared to the reconstruction and physics losses.

\subsection{Physics-aware reconstruction results}
For the following examples, we render synthetic views of a target mesh and use the physics-aware reconstruction framework described above to reconstruct an implicit surface with desirable physical characteristics. We emphasize that during this process, the optimizer has no knowledge of the target mesh topology or 3d positions, i.e, has no strong prior.

\begin{figure}
     \centering
     \includegraphics[width=\linewidth]{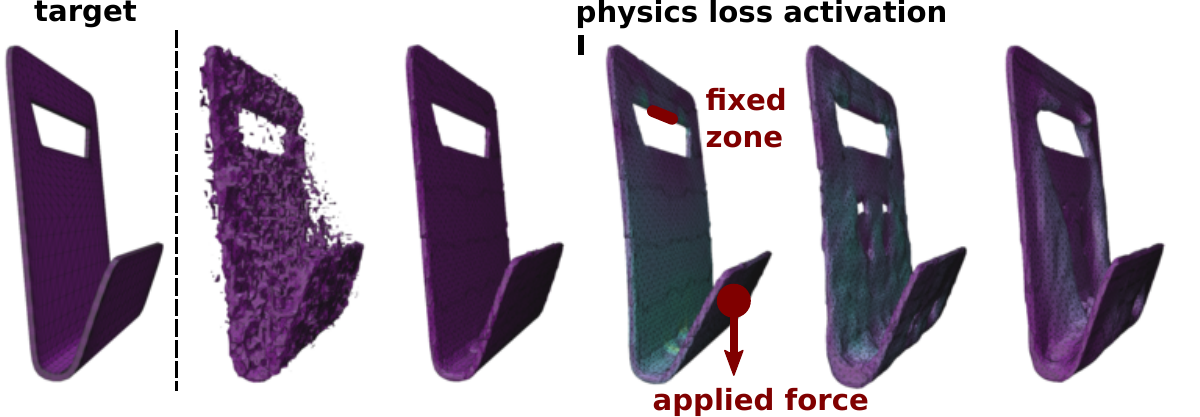}
     \includegraphics[width=\linewidth]{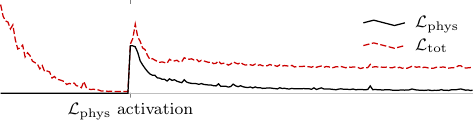}
     \caption{Bracket topology optimized to minimize stress given a predefined load.
     Top: target model (leftmost), then timeline of combined shape reconstruction and stress minimization. The physics-aware loss and prescribed force get activated on the fourth image from the left; shading indicates surface stress intensity.
     Bottom: evolution of the physics and total losses over iterations.
     }
     \label{fig:hook}
 \end{figure}

\paragraph{Stress minimization}
We first apply our method to optimize the shape of an aluminum hook so that stress under some predefined load is minimized (i.e., we use $\phyzloss$ with  $\ell_{\uv} = 0$ and $\ell_{\sigt} = 1$). We use the Stable Neo--Hookean material from \cite{Smith18} with Young Modulus $E_Y=10\text{GPa}$, Poisson ratio $\nu=0.33$, and volumetric mass $\rho=2700\text{kg.m}^{-3}$, and a FlexiCubes grid with resolution 64.
A force of $6$kN is applied to the curved portion of the hook while clamping the top of the slit; see Figure {\ref{fig:hook}}. 
Over the course of the optimization the physics loss $\phyzloss$ is reduced by more than an order of magnitude, with the maximum stress on the surface being similarly reduced.

 \begin{figure*}
        \centering
        \includegraphics[width=\textwidth]{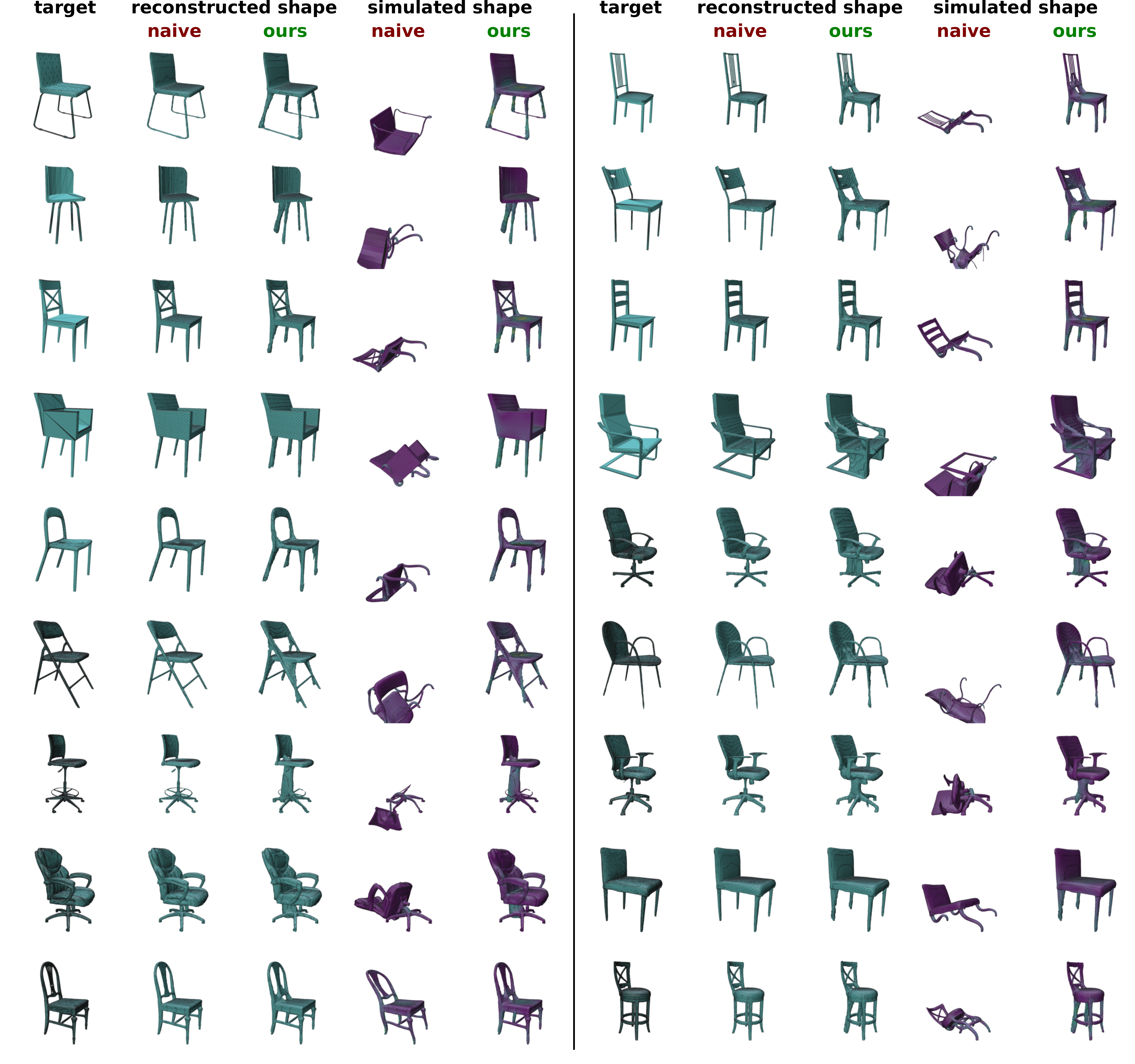}\\
        \caption{
            Physics-aware image-based reconstruction of chair models from the Pix3d dataset such that they can sustain prescribed forces despite being made of a very soft material. For each model, from left to right, target shape, reconstructed shape without (naive) then with (ours) physics-aware loss, simulation of the reconstructed shape without (naive) then with (ours) physics-aware loss.
            We assume a homogeneous  material with density $\rho = 1000kg.m^{-3}$, Young modulus $E_Y = 10$MPa and Poisson ratio $\nu = 0.47$, and apply a downward force of $2.5$kN on the seat and a backward force of $0.5$kN on the backrest.
            On simulation pictures, hue indicates relative stress intensity.
        }
        \label{fig:chairs}
   \end{figure*}

 \begin{figure*}
        \centering
        \includegraphics[width=\textwidth]{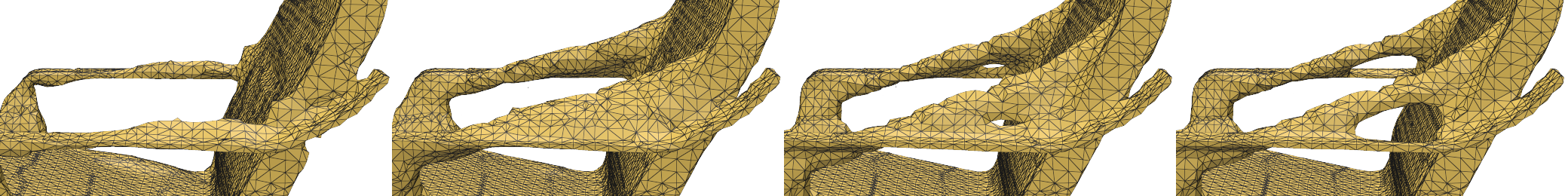}\\
        \includegraphics[width=.24\textwidth]{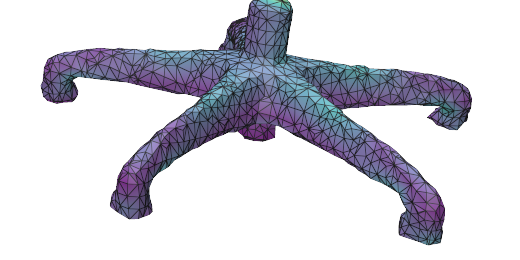}
        \includegraphics[width=.24\textwidth]{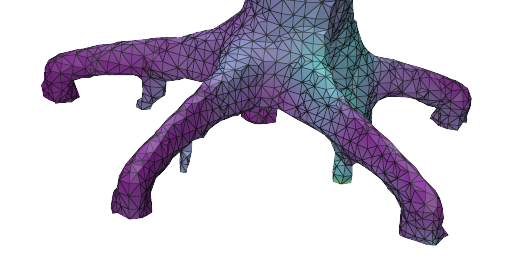}
        \includegraphics[width=.24\textwidth]{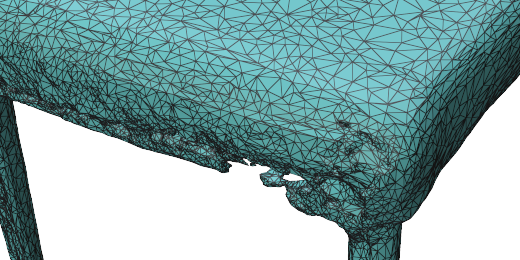}
        \includegraphics[width=.24\textwidth]{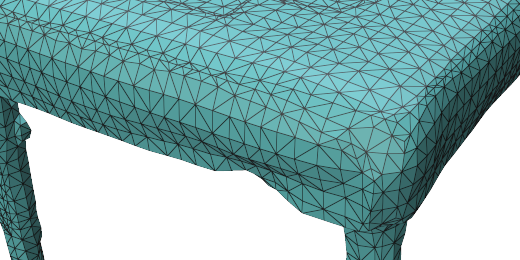}
        \caption{
            Details of some topology changes in our soft chairs example. Top: Additional backrest support being grown then carved out. 
            Bottom, left: Additional feet being grown for support on an office chair. Right: Repairing of a damaged target shape.
        }
        \label{fig:chair-crops}
   \end{figure*}
   
\paragraph{Soft chairs}
To evaluate our method on more challenging material topology and nonlinear effects, 
we select 18 representative chair models from the Pix3D dataset~\citep{pix3d} and equip them with a rubber-like material, with volumetric mass $\rho = 1000\text{kg.m}^{-3}$, Young modulus $E_Y = 10MPa$ and Poisson ratio $\nu = 0.47$. 
Emulating the effect of one person sitting on the chair, we apply a downward force of 2.5kN on the seat and a backward force of 0.5kN on the backrest, with a random perturbation of 10\% of the force direction and point of application at each iteration. Since we do not know in advance where the 3d location of those features, we define our forces in a volumetric fashion over a predefined region of the reconstruction bounding box and scale them according to the actual amount of material in the region. 
The bottom 5\% of each object is kept fixed. We use a timestep $\delt=3$s, loss scaling parameters $\ell_{\uv}=1$ and $\ell_{\sigt}=0.25$, and a FlexiCubes resolution of 64. We run the optimization for 1000 gradient descent iterations, and for each of those run 5 Newton steps of Mixed FEM simulation, which in total takes about 15 to 25 minutes per model (depending on the number of active voxels) on a pair of NVIDIA GeForce RTX 3080Ti GPUs.

The results are depicted in Figure~\ref{fig:chairs}. While applying the forces to the chairs reconstructed without the physics-aware loss usually leads to a complete collapse, the chairs reconstructed with $\phyzloss$ demonstrate much stronger resistance and are easily able to recover their original shape once the perturbations cease being applied. The optimization generally reinforces the chair legs and the seat--backrest junction, but with variations depending on the actual topology, such as the presence (or not) of armrests.
Since we perform reconstruction from images without any strong shape prior --- starting from random SDF values, topological changes are mandatory in the first stages of the optimization. Even once the reconstructed geometry has started to resemble the target,
we keep observing emerging topological changes compared to the mesh used as source; some of which are highlighted in Figure~\ref{fig:chair-crops}.
We emphasize that even in the rare cases where the genus is not modified by the physics loss, the changes to the surface are drastic enough that shape-differentiable simulators working on conforming meshes would require frequent volumetric remeshing~\citep{Huang24, Tozoni21cba}, which our implicit approach avoids entirely.
   
  \begin{figure}
     \centering
     \includegraphics[width=\linewidth]{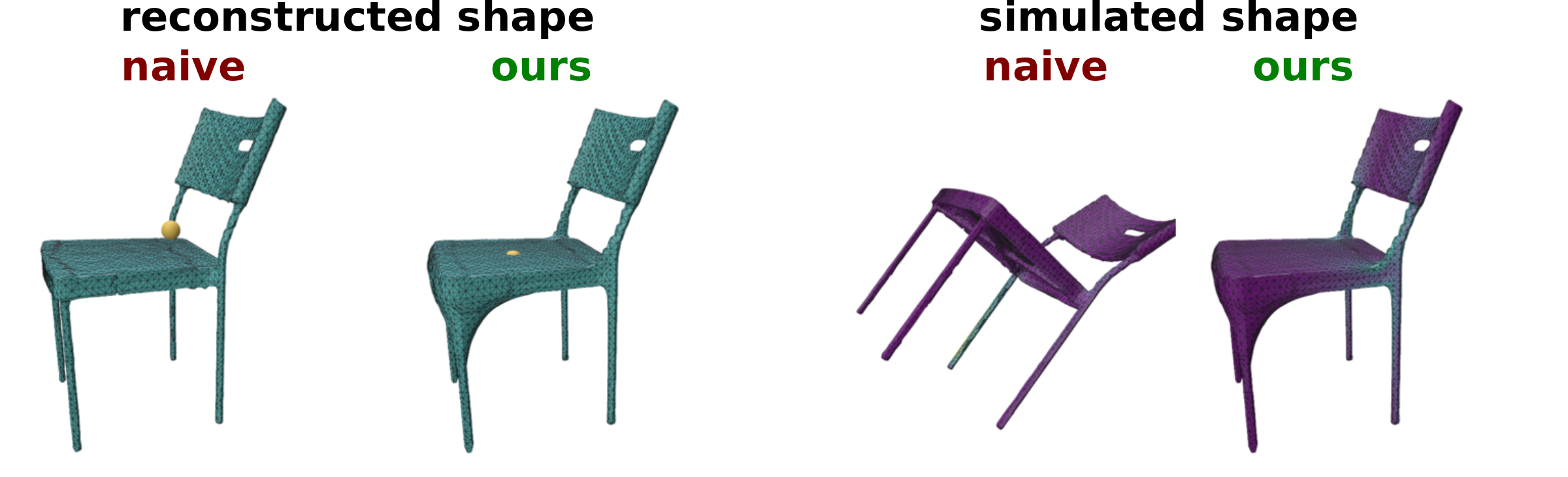}
     \caption{
         Optimizing the stability such that the chair remains stable to a force applied on the backrest. From left to right, reconstructed shape without (naive) then with (ours) physics-aware loss, simulated reconstruction without (naive) then with (ours) physics-aware loss. The yellow ball on the rest geometries indicates the position of the center of mass.
     }
     \label{fig:stability}
\end{figure}

\paragraph{Stability} 
Previously we kept the bottom of the chairs fixed, which is justified given the strong downward applied force. Here, inspired by similar experiments in~\citep{ni2024phyrecon, guo2024physcomp}, we show that our technique can also be leveraged to increase the stability envelope of the reconstructed models. We replace the bilateral clamping with an unilateral constraint modeling the ground--chair contact, and update our Newton loop with an active-set formulation.
We use a much stiffer material so that the chair behaves rigidly ($E=100$GPa), apply a downward-and-backward-pointing force on the backrest, and pick a model that looks propitious to toppling. Figure~\ref{fig:stability} shows that the physics-aware loss will add material to the front of the chair such that the center of mass moves forward and resists the applied push.

 \begin{figure}
      \centering
      \includegraphics[width=\linewidth]{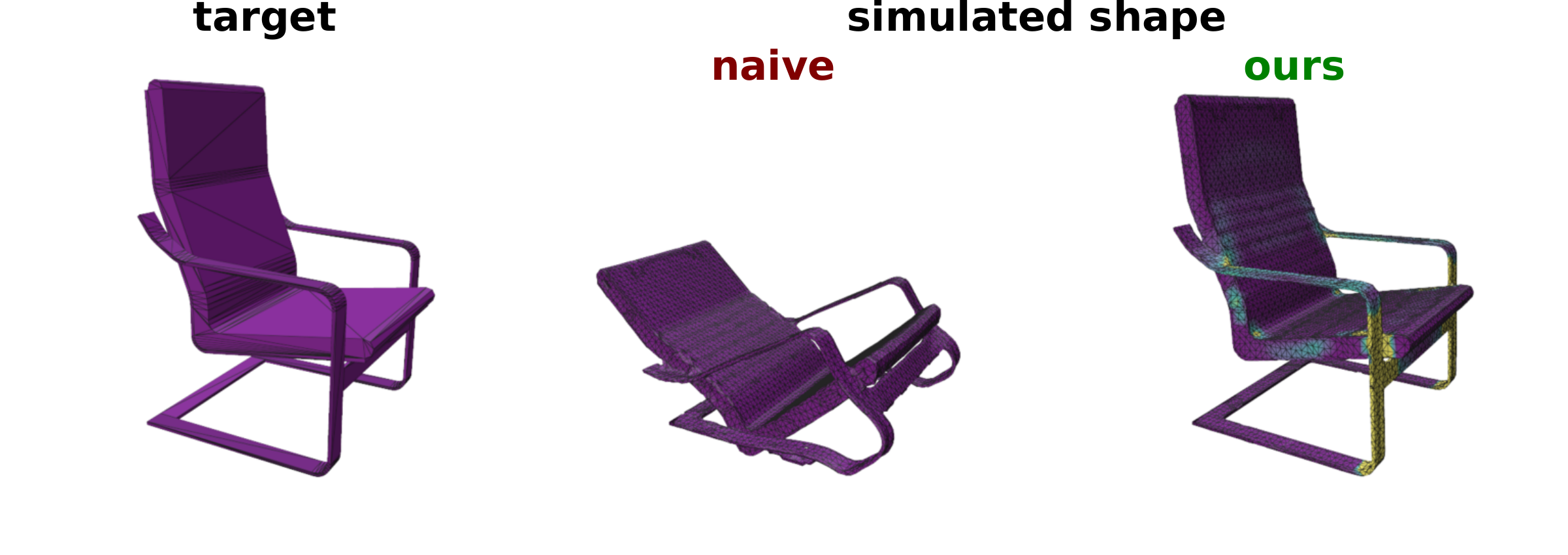}
      \caption{
          Concurrent optimization of the shape and Young modulus of the chair. From left to right, target shape, simulated reconstructions without (naive) and with (ours) physics-aware loss. Shading indicates regions that are made stiffer.
      }
      \label{fig:stiffopt}
  \end{figure}
  
\paragraph{Material optimization}
Up until now we only allowed the optimizer to modify the shape of the model, keeping the material homogeneous; here we also allow modification of the Young Modulus. This makes the problem somewhat easier, as now the physics loss and reconstruction loss can act on orthogonal parameters. In our framework, we can just set $\ell_{\uv}$ and $\ell_{\sigt}$ to small values so that the optimizer will favor $\FCloss$ over $\phyzloss$ for the shape parameters.
Figure~\ref{fig:stiffopt} shows the result of this process under the constraint that the Young modulus should not be increased more than $10^4\times$ and with additional $L_1$ regularization of the stiffening parameter. Unsurprisingly, the legs of the chair and the junction between legs and seat are the regions that the optimizer prioritize for stiffening.

 \begin{figure}
     \centering
     \includegraphics[width=\linewidth]{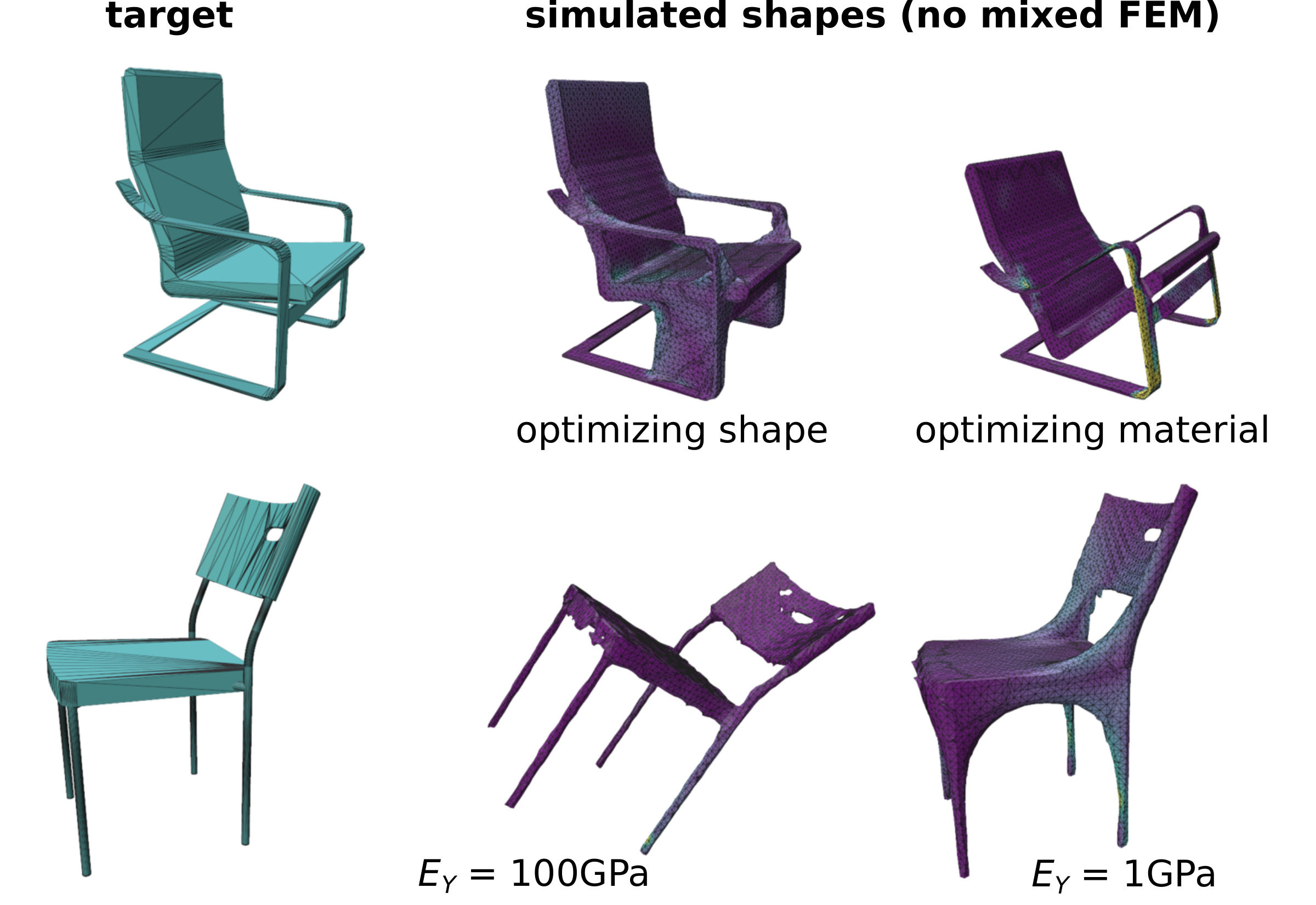}
     \caption{
     Results using displacement-only FEM instead of our Mixed FEM. 
     Top: from left to right: target shape, simulation of shape reconstructed with topology optimization, simulation of shape reconstructed with material stiffness optimization.
     Bottom: stability optimization, from left to right: target shape, simulation of shape reconstructed with $E_Y=100$GPa, then with $E_Y=1$GPa.
     }
     \label{fig:nomfem}
 \end{figure}

 \begin{figure}
     \centering
     \includegraphics[width=\linewidth]{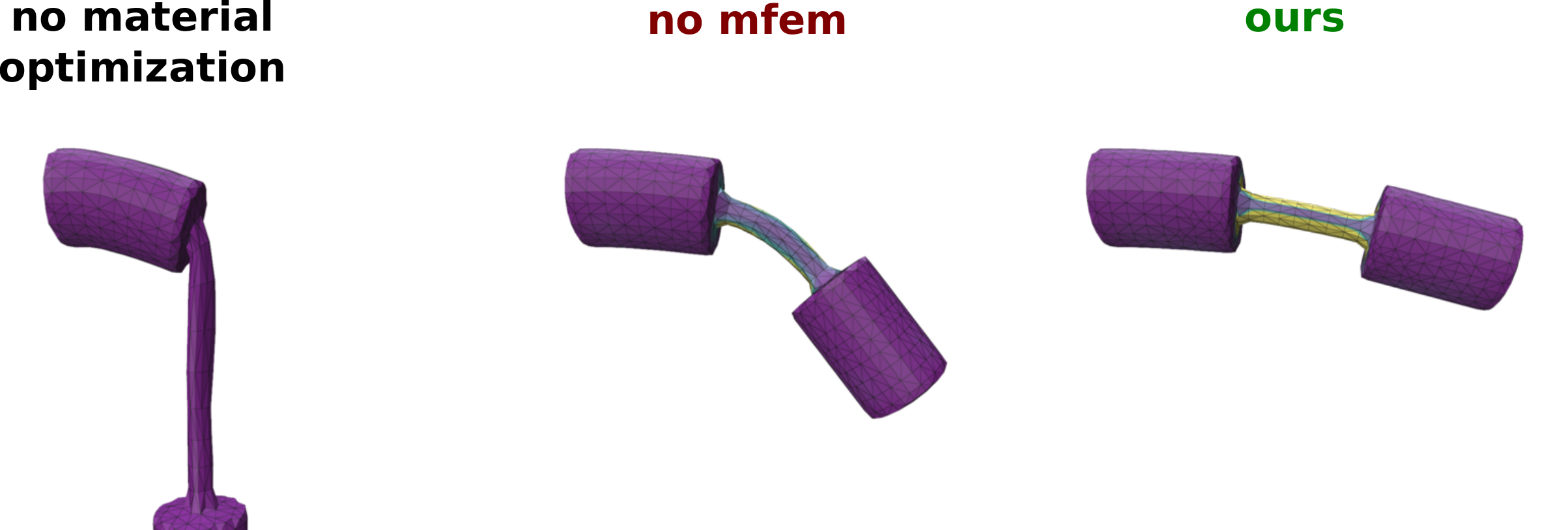}
     \caption{
Simulation of the reconstructed elastic dumbbell from Section~\ref{sec:dumbbell} without material optimization~(left), with material optimization and displacement-only FEM~(middle), with material optimization and our mixed FEM~(right). Hue shows material stiffness scaling, from $1\times$ (purple) to $250 \times$ (yellow). Slower convergence of displacement-only FEM with high stiffness ratios causes the optimizer to underestimate the physics loss $\phyzloss$, leading to insufficient stiffening and more sagging when re-simulating the reconstructed object.
     }
     \label{fig:dumbbell-stiffopt}
 \end{figure}

\begin{figure}
     \centering
     \includegraphics[width=\linewidth]{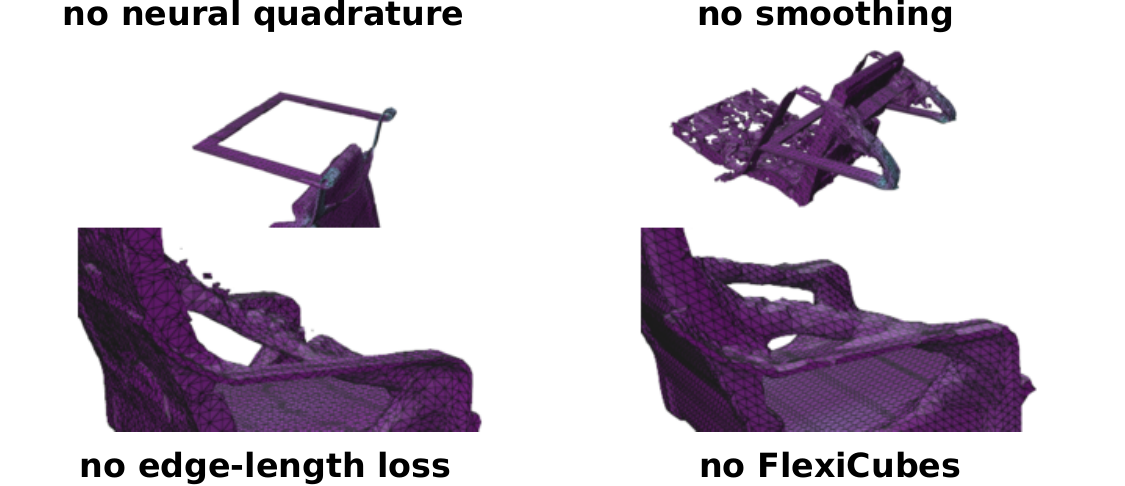}
     \caption{
     Simulation of reconstructed shapes with various parts of our framework removed: top left: no Neural quadrature; top right: no smoothing preconditioner; bottom left: no edge-length loss; bottom-right: using dual marching-cube instead of FlexiCubes (cropped details).
     }
     \label{fig:ablation}
 \end{figure}

\paragraph{Ablation studies}
Having demonstrated the physics-aware reconstruction capabilities of our framework, we proceed to study the importance of its individual components, and show results in Figure~\ref{fig:ablation}.
First, replacing the neural quadrature with Full or Clip quadrature formulas hinder convergence entirely. Indeed,
the gradient of $\phyzloss$ with respect to the vertex SDF values becomes zero, so that the optimizer will only move the vertex positions, whose motion is constrained by the FlexiCubes parameterization to stay under one voxel-size. 
Next, when removing the smoothing preconditioner, the optimizer does manage to reduce the physics loss, but the reconstructed surface is hardly usable.
Removing the edge length loss $\edgelenloss$ makes the reconstructed surfaces bumpier, and tend to produce floaters. 
Finally, sharp features are no longer well captured when limiting the optimized parameters to the vertex SDF values, i.e, falling back to a standard dual marching-cube.

We also show in Figure~\ref{fig:nomfem} that our method keeps working when using displacement-only FEM rather than Mixed FEM, and thus, should be compatible with other differentiable simulators that support hexahedral elements~\cite[e.g,][]{Huang24}; but with degraded robustness. For our initial soft chair optimization problem, classic FEM yields a reconstruction similar to the Mixed FEM case from Figure~\ref{fig:chairs}. For the material optimization experiment the results are still reasonable, but, due to the slower convergence of displacement-only FEM with high stiffness ratios, the physics loss is underestimated and the reconstructed model is subject to more sagging than in the Mixed FEM version from Figure~\ref{fig:stiffopt}; see also Figure~\ref{fig:dumbbell-stiffopt} for a simpler variant of this experiment.
Results are worst for the stability optimization example; here, we need to reduce the stiffness by two orders of magnitude to obtain a stable reconstruction.

\begin{figure}
     \centering
     \includegraphics[width=\linewidth]{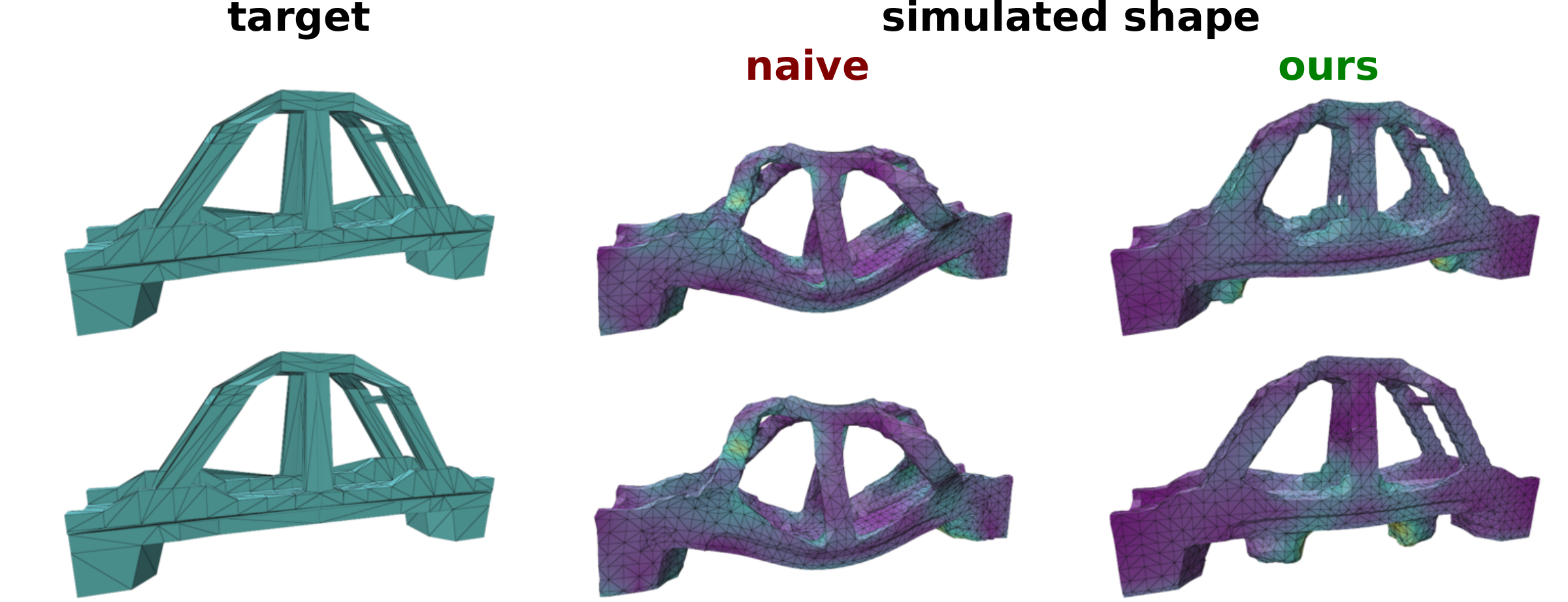}
     \caption{From left to right: target shape and simulation of the shapes reconstructed without (naive) and with (ours) physics-aware loss. Top: using with tri-quadratic elements and the order-4 ``neural'' quadrature; bottom: using trilinear elements and the order-2 ``neural'' quadrature.}
     \label{fig:bridge}
\end{figure}

\paragraph{Quadratic elements}
We verify that our method also works with higher-order elements.  Figure~\ref{fig:bridge} demonstrates optimization of an elastic bridge model under prescribed load on a $48^3$ grid, using tri-quadratic displacements and the order-4  27-points learned quadrature. As the resolution of the FlexiCubes grid is already high enough to resolve thin features of the target shape, the higher-order result remains qualitatively similar to the trilinear version.

\begin{figure}
     \centering
     \includegraphics[width=\linewidth]{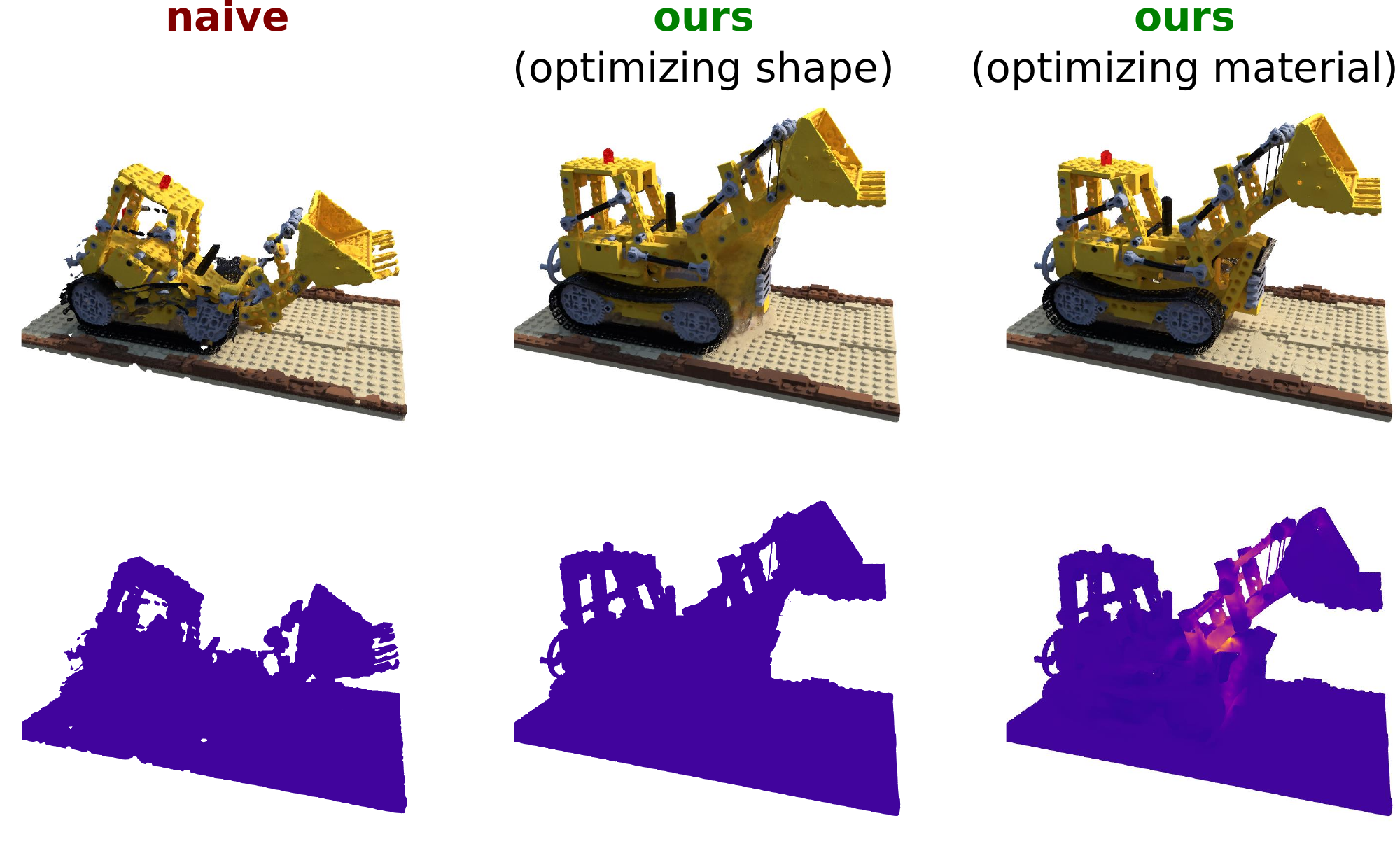}\\
     \caption{
  Physics-aware multiview reconstruction of the Bulldozer scene from NeRF synthetic dataset. Simulation results from left to right: initial shape before \( L_{\text{phys}} \) is applied, optimizing geometry, optimizing material parameters. Shading indicates regions that are made stiffer.
     }
     \label{fig:bulldozer}
 \end{figure}

\paragraph{Physics-aware photogrammetry}
Our physics-aware shape reconstruction formulation may also be integrated into more complex photogrammetry pipelines to allow for the joint optimization of shape, lighting, and both physical and rendering materials. We leverage \textsc{nvdiffrec} from \citet{Munkberg_2022_CVPR}, which supports FlexiCubes as its geometry representation. 
For this example we use again the setup described in Section~\ref{sec:losses}; this time, instead of defining our rendering target as random synthetic views of a known mesh, we use a fixed list of 100 images and corresponding camera transforms from the NeRF dataset~\citep{mildenhall2020nerf}. We first optimize the rendering loss \( L_{\text{render}} \) from the \textsc{nvdiffrec} pipeline without modification. After the shape starts converging, we begin blending in the physics loss function \( L_{\text{phys}} \). We can again optimize both the shape and material parameters of our reconstructed object to minimize sagging under a prescribed force (Figure~\ref{fig:bulldozer}).

\section{Limitations and Future Work}
While our neural quadrature rule can compute integrals over the possibly complex material domain induced by the SDF, the displacement degrees of freedom are still those of the underlying continuous shape functions. As such, even if the SDF defines two disconnected material regions within a given voxel, the simulation won't allow them to separate arbitrarily. 
This is detrimental for our application as this means that small disconnected pieces of material (``floaters``) adjacent to the bulk of the object will tend to stick to the surface instead of falling down --- meaning that they will have a low displacement loss  
and the optimizer won't be eager to prune them. While our edge-length loss helps reduce this phenomenon, other formulations such as the perimeter regularizer from \citet{Maestre23toros} would be worth investigating. A possibly more satisfying solution that we intend to explore in future work is the of addition of new degrees of freedom to the disconnected voxel configurations, in the vein of XFEM~\citep{Moes99, Koschier17} or CPIC~\citep{Hu18}. 
Additionally, we do not consider evaluation of integrals over the boundary of the domain, as these are not needed for the presented tasks. 
If desired, isosurface meshes can easily be extracted and embedded for integration.

Another limitation of our physics-aware reconstruction algorithm is the lack of global convergence, meaning that differences in the initial random SDF values lead to variations in the final optimized shape. The end result is also very sensitive to the choice of loss
function; exploring the definition of more perceptual losses would be an interesting area of research.
Our current reconstruction speed is also not yet suitable for interactive applications; we would like to bridge this gap in the future. 

In future work, we also want to explore whether other architectures could improve the accuracy and efficiency of our technique. 
While we have focused on solid elasticity in this work, we also want to take advantage of the analogy between quadrature points and Particle-in-Cell integration to investigate differentiable initialization of MPM simulations from implicit surfaces. Combined with particle-based  techniques like PAC-NeRF~\citep{li2023pac}, this would allow computing derivatives of the simulation end-state with respect to the material occupancy function, with applications to single-pass reconstruction, identification and shape optimization of plastic materials.

\section{Conclusion}
We have presented a neural integration technique that improves the quality of voxel-based implicit volume simulations at negligible additional runtime cost, and shown how to combine it with a Mixed FEM solver to efficiently perform elasticity simulation on continuously evolving domains.
Our technique allows straightforward differentiation of the simulation results with respect to the implicit volume parameters, making itself particularly suitable for topology optimization tasks, and providing a first foray into physics-enabled shape reconstruction.

\bibliographystyle{ACM-Reference-Format}
\bibliography{flexisim}

\appendix

\section{Neural Quadrature Training and evaluation}
\label{app:training}

For a quadrature of target order $d$, we use $n_Q:= (d/2 + 1)^3$ quadrature points per voxel and a  test polynomial basis $\testbasis$ of cardinality $n_P := (d+1)^3$, chosen as the 3D tensor product of 1D Lagrange polynomials with Lobatto--Gauss--Legendre nodes $(x^{\text{LGL}}_i)$,
\begin{align*}
    \testbasis &:= \left\{ P_{ijk} ,  0 \leq i, j, k \leq d \right\}, \\
    P_{ijk}(x, y, z) &:= P^{\text{LGL}}_i(x)P^{\text{LGL}}_j(y)P^{\text{LGL}}_k(z), \\
    P^{\text{LGL}}_i(x) &:= \frac{ \prod_{l \neq i} \left(x - x^{\text{LGL}}_l\right) }{ \prod_{l \neq i} \left(x^{\text{LGL}}_i - x^{\text{LGL}}_l\right)}.
\end{align*}
We note $n_\varphi := 2^3$ the number of input SDF values per voxel,  and $(\trilinear_j)$ the $n_\varphi$ corresponding trilinear interpolation functions.

Finally, we denote by $n_B$ the batch size, i.e, the number of voxels being simultaneously evaluated.

\begin{figure}
    \centering
    \includegraphics[width=\linewidth]{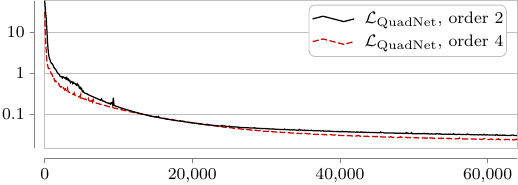}
    \caption{Convergence of MLP training over iterations}
    \label{fig:mlp-conv}
\end{figure}

\begin{figure}
    \centering
    \includegraphics[width=.85\linewidth]{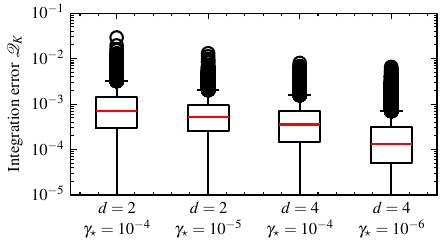}
    \includegraphics[width=.85\linewidth]{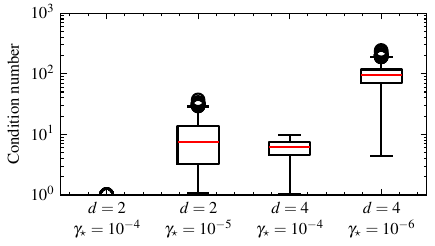}
    \caption{Statistics of integration error and conditioning $\max{w_j}/\min{w_j}$ over 1000 random voxels for networks trained with order $d$ and conditioning loss scaling factor $\gamma_\star$.}
    \label{fig:mlp-comp}
\end{figure}

We reiterate that as our network works independently for each voxel, we do not need to construct a dataset containing macroscopic shapes: for our training data, we simply generate a set of $2^{24} \times 8$ random values sampled from a normal distribution. 

For each voxel, the $n_{\varphi}=8$ corner values do not need to represent a proper SDF satisfying the eikonal equation; we actually want the network to be robust to improper SDFs, and the first Normalization layer of our network (Algorithm~\ref{alg:normalize}) will remap the input such that the gradient at the voxel center is unitary. 
This Normalization layer is followed by a standard 5-layers-deep MLP with ReLu activations, then a final Remapping layer (Algorithm~\ref{alg:remap}) yielding the final quadrature points and weights for each voxel.

For each randomly generated voxel, we generate the ground truth data through brute-force integration at resolution $32^3$ of all Lagrange polynomials of our chosen basis $\testbasis$ multiplied by the indicator function of the SDF interior $\varphi < 0$ (Algorithm~\ref{alg:ground_truth}). This ground truth is compared to the integrals computed using the inferred quadrature points (Algorithm~\ref{alg:quad_eval}) as part of our overall training loss (Algorithm~\ref{alg:loss}), which also incorporates penalties for points drifting outside the voxel and large weight ratios.

We additionally generate at test set of $2^{10}$ distinct random voxels and associated ground truth integral values, and periodically evaluate the loss function and those as the network is training; resulting curves are shown in Figure~\ref{fig:mlp-conv}.

We choose the training batch size for the AdamW optimizer as the highest that can fit in our GPU memory; in our case, $n_B = 2^{18}$. Once training is finished, we evaluate the quality of our network using yet another validation set of random voxels and ground truth integrals pair --- this time, with random values sampled according to a uniform distribution. We visualize separately the integration error and conditioning loss for different order and values of the conditioning penalty coefficient $\gamma_\star$ in Figure~\ref{fig:mlp-comp}. Generally, we want to pick $\gamma_\star$ such that it yields a reasonable conditioning but not at the detriment of integration accuracy; in our examples, we use networks trained with $\gamma_\star = 10^{-5}$ for order-2 and $\gamma_\star = 10^{-6}$ for order-4 network.

\begin{algorithm}[t]
\SetAlgoNoLine
\KwIn{
$\varphi$: Tensor of voxel corner SDF values, size $n_B \times n_\varphi$;
}
\KwOut{
$\Upsilon^{\text{GT}}$: ground-truth integral values, size $n_B \times n_P$.
}
\KwParameters{$n_{\text{GT}}:$ resolution of brute-force integration}

$\Upsilon^{\text{GT}} = 0$\;
$h := 1 / n_{\text{GT}}$\;
\ForEach{voxel $b$ in batch}{
    \For{$0 \leq i, j, k < n_{\text{GT}}$}
    {
        \tcp{Interpolate at uniformly sampled locations}
        $x, y, z := (i+0.5)h, (j+0.5)h, (k+0.5)h$ \;
        $\varphi_{xyz} := \sum_{l < n_\varphi} { \varphi_l \trilinear_l(x, y, z) }$ \;

        \If(\tcp*[h]{In SDF interior}){$\varphi_{xyz} < 0.0$} {
            \ForEach{polynomial $P_l$ in $\testbasis$}{
                $\Upsilon^{\text{GT}}_{b,l} = \Upsilon^{\text{GT}}_{b,l} + P_l(x, y, z) h^3$
            }
        }
    }
}

\caption{Ground truth integrals generation}
\label{alg:ground_truth}
\end{algorithm}

\begin{algorithm}[t]
\SetAlgoNoLine
\KwIn{
$\varphi$: Tensor of voxel corner SDF values, size $n_B \times n_\varphi$;
}
\KwOut{
$\widebar{\varphi}$: MLP output tensor, size $n_B \times n_\varphi$.
}

\ForEach{voxel $b$ in batch}{
    \tcp{Normalize SDF gradient at voxel center}    
    $\gv_b := \sum_{j<n_\varphi} \varphi_{b,j} \nabla \trilinear_j(0.5, 0.5, 0.5)$  \;
    $\widebar{\varphi_b} = \varphi_b / \left( \Vert \gv_b \Vert + 10^{-8} \right) $ 

    \tcp{Shift full and empty voxels closer to origin}
    \uIf{$\min_j \varphi_{b,j} > 1$} {
        $\widebar{\varphi_b} = \widebar{\varphi_b} - \min_j \varphi_{b,j} + 1$\;
    }
    \uIf{$\max_j \varphi_{b,j} < - 1$} {
        $\widebar{\varphi_b} = \widebar{\varphi_b} - \max_j \varphi_{b,j} - 1$\;
    }
}   

\caption{MLP input normalization layer}
\label{alg:normalize}
\end{algorithm}

\begin{algorithm}[t]
\SetAlgoNoLine
\KwIn{
$\xv$: MLP output tensor, size $n_B \times n_Q \times 4$;
}
\KwOut{
$\yv, \wv$: tensors of quadrature point coordinates and weights, size $n_B \times n_Q \times 3$ and $n_B \times n_Q$.
}
\KwData{$\yv^{\text{GL}}, w^{\text{GL}}$: 3D Tensor product of 1D Gauss--Legendre points and weights of order $d$, size $n_Q \times 3, n_Q$}
$\yv_{\cdot} = \yv^{\text{GL}} + \tanh \left( \xv_{\cdot, \cdot, 0\ldots 2} \right)$  \;
$w_{\cdot} = w^{\text{GL}} \exp \left( \xv_{\cdot, \cdot, 3} \right)$ \;
\caption{MLP output remapping layer}
\label{alg:remap}
\end{algorithm}

\begin{algorithm}[t]
\SetAlgoNoLine
\KwIn{
$\yv$: tensor of quadrature point coordinates, size $n_B \times n_Q \times 3$;
$w$: tensor of quadrature point weights, size $n_B \times n_Q$.
}
\KwOut{Tensor $\Upsilon$ of integral values for all Lagrange polynomials in the test basis $\testbasis$, size $n_B \times n_P$.}
\ForEach{polynomial $P_l$ in $\testbasis$}{
    $\Upsilon_{\cdot,l} = \sum_{q < n_Q} w_{\cdot, q} P_l(\yv_{\cdot, q})$
}
\caption{Quadrature evaluation layer}
\label{alg:quad_eval}
\end{algorithm}

\begin{algorithm}[t]
\SetAlgoNoLine
\KwIn{
$\yv$: tensor of quadrature point coordinates, size $n_B \times n_Q \times 3$;
$w$: tensor of quadrature point weights, size $n_B \times n_Q$;
$\Upsilon^{\text{GT}}$: ground-truth integral values, size $n_B \times n_P$.
}
\KwOut{$\genloss_{\textrm{QuadNet}}$: scalar loss.}
\KwParameters{
$\gamma_\square$: scaling factor for quadrature point interior loss;
$\gamma_\star$: scalar factor for the conditioning loss
}
$\Upsilon := \textrm{EvalQuadrature}(\yv, \wv)$ \tcp*{Algorithm~\ref{alg:quad_eval}}
$\mathcal{Q}_K := \Vert \Upsilon - \Upsilon^{\text{GT}} \Vert_2^2$\;
$\mathcal{Q}_{\square} := \left\Vert \yv - \text{clamp}(\yv, min=0, max=1) \right\Vert_2 $ \;
$\mathcal{Q}_{\star} := \left\Vert \log(\max_{q < n_Q} w_{\cdot, q}) - \log(\min_{q < n_Q} w_{\cdot, q}) \right\Vert_1 $ \;
$\genloss_{\textrm{QuadNet}} := \mathcal{Q}_K + \gamma_\square \mathcal{Q}_\square + \gamma_\star \mathcal{Q}_\star$ \;
\caption{Loss layer}
\label{alg:loss}
\end{algorithm}

\begin{figure}
    \includegraphics[width=\columnwidth]{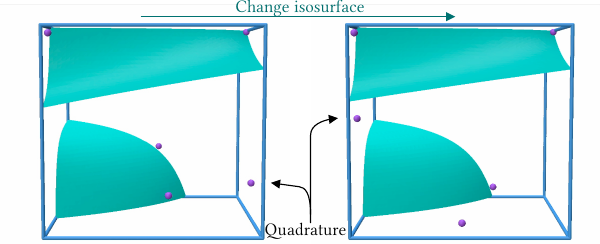}
    \caption{Using a linear program to select a sparse subset of quadrature points can lead to large jumps in quadrature point position for small changes in isovalues.}
    \label{fig:longva}
\end{figure}

An advantage of our neural quadrature scheme is that it produces, by construction, smooth, differentiable output. Sparse subset selection methods via linear programming~\citep{Longva20} produce accurate, efficient quadrature schemes but the under constrained nature of the subset selection problem, and the lack of additional regularizers, lead to large jumps in quadrature point location as a function of isosurface value (Figure~\ref{fig:longva}), making them unsuitable for gradient-based optimization. 

\begin{figure}
    \centering
    \includegraphics[width=.85\linewidth]{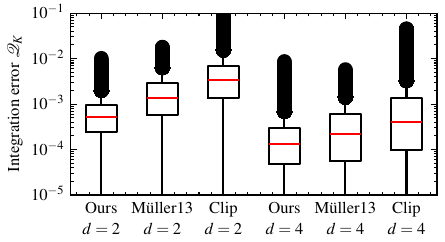}
    \caption{Statistics of integration error over 1000 random voxels for our neural quadrature, the weight-only moment-fitting technique from~\citet{Muller13}, and the Clip quadrature at order $d$.}
    \label{fig:mlp-comp-muller-clip}
\end{figure}

Figure~\ref{fig:mlp-comp-muller-clip} shows that our neural integration rule is also significantly more accurate than the Clip quadrature of the same degree, and more accurate than the weight-only moment-fitting approach from \citet{Muller13} when using the same number of quadrature points, though the gap reduces with integration order. Note that the conditioning number of the Clip a quadrature will be infinite for any partially filled voxel, and that the method from \citet{Muller13} may produce null or negative weights,
while our neural approach ensures positivity and reasonable weight ratios. 
On a NVIDIA GeForce RTX 3080Ti GPU, computing the order-4 quadrature points for $2^18$ voxels takes 6ms for the Clip quadrature; 16ms for our neural quadrature; and more than 1s for the technique from~\cite{Muller13}, which requires evaluating the target moments using high-resolution integration, here with $2^12$ samples per voxel.

\section{Four-field Mixed FEM implementation}
\label{app:mfem}

\subsection{Newton optimizer}
\label{sec:mfem_continuous}
We proceed to solve system~(\ref{eq:nl_cm}--\ref{eq:nl_cst}) using projected Newton iterations to compute the step direction $(\delta \uv, \delta \St, \delta \Rt, \delta \sigt)$, with 
$\delta \Rt \in \Skews$ such that $\Rt^{k+1} = \Rt^k + \Rt^k \delta \Rt$. 
Linearizing the residual around the current iterate $(\uv^k, \St^k, \Rt^k, \sigt^k)$ yields the 
linear forms
\begin{align*}
\psi^k_{,\St} &:= \psi_{,\St}(\St^k; \cdot), &
c^k &:= c(\uv^k, \St^k, \Rt^k, \cdot),
\end{align*}
and the bilinear forms 
\begin{align*}
c^k_{,\St} &:= c_{,\St}(\Rt^k; \cdot, \cdot), \\
c^k_{,\Rt}(\delta \Rt, \lambdat) &:= \int_{\Omega}{\Rt^k \delta \Rt \St^k : \lambdat^T},\\
h^k(\delta \St, \taut) &:= \int_{\Omega} \delta \St : \Pi\left(\pddX{\Psi}{\St} \left(\St^k\right) \right) :  \taut + \varepsilon \int_{\Omega} \delta \St:\taut, \text{ and }\\
\epsilon^k(\delta \Rt, \omegat) &:= \varepsilon \int_{\Omega} (\delta \Rt \St^k):(\omegat \St^k),
\end{align*}
where the $\Pi$ operator removes negative eigenvalues from the Hessian of $\Psi$. 
At each Newton iteration, we thus solve
\begin{equation}
    \label{eq:newton_iter}
    \begin{aligned}
    a(\uv^k + \delta \uv, \vv) + c_{,\uv}(\vv, \sigt^k + \delta \sigt) &= b(\vv) & \forall \vv &\in \Vs, \\
    h^k(\delta \St, \tau) + c_{,\St}^k(\taut, \sigt^k + \delta \sigt + \varepsilon \Ct^k) &= -\psi^k_{,\St}(\taut)  &\forall \taut &\in \Syms,\\
    \epsilon(\delta \Rt, \omegat) + c_{,\Rt}^k(\omegat, \sigt^k + \delta \sigt + \varepsilon \Ct^k) &= 0 &\forall \omegat &\in \Skews,\\
    c_{,\uv}(\delta \uv, \lambdat) + c_{,\St}^k(\delta \St, \lambdat) + c_{,\Rt}^k(\delta \Rt, \lambdat) &= -c^k(\lambdat) &\forall \lambdat &\in \Ts,\\
    \end{aligned}
\end{equation}
with $\Ct^k$ the current constraint residual, $\Ct^k := \Ct(\uv^k, \St^k, \Rt^k)$. 

We equip our Newton loop with a backtracking line-search; however, as we are optimizing under equality constraints, iterates will not remain perfectly feasible and thus we can't directly use the incremental potential~(\ref{eq:ip}) as our objective function. Instead we adopt the merit function $$\phi(\uv, \St, \Rt) := \frac 1 2 a(\uv, \uv) - b(\uv) +\psi(\St) + \int_{\Omega} E_Y \Vert C(\uv, \Rt, \St) \Vert,$$ combined with Armijo's acceptance rule~\cite{Nocedal06}.

\paragraph{Rotation regularization}
In practice, we find that the robustness of convergence can be increased by augmenting the penalization term $\varepsilon$ in the bilinear form $\epsilon^k$ with an additional rotation regularization coefficient $\vartheta > 0$. Indeed, we want the skew-symmetric update $\delta \Rt$ to remain small enough that it still represents a valid rotation increment. In our examples we choose $\vartheta$ equal to the residual rotational stress norm, $\vartheta = \Vert \Rt^T \sigma - \sigma^T \Rt \Vert$. Note that this additional regularization term does not change the problem solution. We do not add the $\vartheta$ regularization term when assembling the linear system corresponding to the backward step in the simulation adjoint computation.

\subsection{Discrete Mixed Elements}
\label{app:discrete_mfem}

After choosing discrete basis for our element spaces and performing numerical integration for all the terms of Equation~(\ref{eq:newton_iter}), computing the Newton step direction amounts to solving the linear system
\begin{equation}
    \label{eq:newton_sys}
\left[\begin{array}{cccc}
     A & & & C_{,\uv}^T \\
     & H^k & & C_{,\St}^{k,T} \\
     & & E & C_{,\Rt}^{k,T} \\
     C_{,\uv} & C^k_{,\St} & C^k_{,\Rt} &
\end{array} \right] \left( \begin{array}{c}
     \delta \uv \\
     \delta \St \\
     \delta \Rt \\
     \delta \sigt \\
\end{array} \right) = \left( \begin{array}{r}
\bv - A\uv^k - C_{,\uv}^{T} \sigt^k \\ 
-\psiv^k - C_{,\St}^{k,T} \left(\sigt^k + \varepsilon \cv^k\right) \\ 
- C_{,\Rt}^{k,T} \left(\sigt^k + \varepsilon \cv^k\right) \\ 
- \cv^k
\end{array} \right).
\end{equation}

While the linear system will always have this general shape whatever our choice of discrete spaces, the latter will impact the sparsity pattern of the matrices and our options for solving it.
As is standard in finite-element elasticity, we assume that our basis functions $(N_i)_K$ over each mesh element $K$ are Lagrange polynomials defined over a set of nodes $(\xv_i)_K$, meaning that $N_i(\xv_j) = \delta^i_j$. Due to different continuity requirements, we use distinct basis functions and nodes for the displacement space $\Vs$ and the tensor spaces $\Ts$, $\SOs$, $\Syms$ and $\Skews$, as explained below.

For the displacement space $\Vs$ we need $H^1$-compatible elements, i.e, continuity of the basis functions across neighboring elements. We restrict our choice to usual $P_d$ Lagrange elements or so-called ``serendipity'' $S_d$ elements that do not contain interior nodes. For $P_1$ and $S_1$ this means usual trilinear shape functions with one node at each grid vertex; for $S_2$, one node at each vertex and one node at the middle of each edge; etc.

For the $\Ts$, $\SOs$, $\Syms$ and $\Skews$ spaces however, we only need to discretize $\Lts$ (our mixed formulation does not require evaluating the derivatives of the stress or strain fields). Continuity across elements is not required and we get the liberty to locate the tensor field nodes $(\xv_i)_K$ anywhere within $K$. To get an insight about how to pick their positions, we look once again at numerical integration. To integrate a function $f$ on element $K$ we resort to a discrete quadrature formula with weights $(w_p)_K$ and evaluation points $(\yv_p)_K$, that is, 
$$
\int_K {f} \sim \sum_p {w_p f(\yv_p)}.
$$
As discussed in Section~\ref{sec:optim_qp}, the weights and points are typically picked such that the formula is exact for polynomials up to a given order. Now when evaluating the matrix $A$ for a bilinear form defined from two sets of basis functions $(N_i)_K$ and $(N_j)_K$, we get 
$$
A_{i,j} := \int_K {N_i N_j f} \sim \sum_p {w_p N_i(\yv_p) N_j(\yv_p) f(\yv_P)},
$$
meaning that if we pick our nodes and quadrature points such that $(\xv_i)_K = (\xv_j)_K = (\yv_p)$, then $A_{i,j} = \sum_p w_p \delta_i^j f(\yv_p)$, i.e,
the matrix $A$ becomes block diagonal. 
For a discontinuous Lagrange polynomial basis of chosen degree $d$, we thus pick the nodes $(\xv_i)_K$ to correspond to the points of a quadrature formula that maximizes the order of accuracy for this number of points. In particular for quadrilateral or hexahedral elements, this is the usual Gauss--Legendre points, which yield a quadrature formula that is exact for polynomials up to order $2d$. For triangle and tetrahedral elements, we get quadrature formulas of order $2d$ for $d\leq 1$, while for for higher degree polynomials we rely on numerical optimization.

By using this choice of quadrature points to define the Lagrange nodes for the 
discrete subspaces $\Ts$, $\SOs$, $\Syms$ and $\Skews$, we thus render the matrices $H^k$, $E$, $C_{,\St}^k$ and $C_{,\Rt}^k$ block-diagonal, while keeping a good order of accuracy for the numerical integration.

To finalize the Mixed FEM discretization, it remains to relate the displacement and tensor spaces. For hexahedral elements, we use serendipity elements $S_d$ of degree $d$  for the displacement and tensor products of element-wise discontinuous Lagrange polynomials (with Gauss--Legendre nodes) of the same degree $d$ for tensors.
For tetrahedral elements, we use continuous Lagrange polynomials $P_d$ of degree $d$  for the displacements and discontinuous Lagrange polynomials of degree $d-1$ for the tensor spaces.
This means that for hexahedral elements, the quadrature formula defined by the tensor nodes will be enough to integrate accurately all bilinear forms, while for tetrahedral elements we will need to use a distinct, higher-order quadrature formula for the displacement forms.

\subsection{Solving the linear system}
While it is possible to directly solve Equation~{\ref{eq:newton_sys}} using a saddle-point solver, in practice we find it more efficient to follow ~\citet{Trusty22} and perform double condensation. 
Here for the sake of clarity we focus on the current Newton iteration and drop the $k$ index for matrices and vectors.

First, thanks to the non-zero Augmented--Lagrangian penalization coefficient $\varepsilon$ and the semi-definite projection of the elasticity Hessian, $H$ and $E$ are positive definite (as long as $S$ in non-singular). As they are also block-diagonal, they are easily invertible, 
and we can eliminate the $\delta \St$ and $\delta \Rt$ unknowns to obtain

\begin{equation*}
\left[\begin{array}{cc}
     A &  C_{,\uv}^T \\
     C_{,\uv} & -\Lambda 
\end{array} \right] \left( \begin{array}{c}
     \delta \uv \\
     \delta \sigt \\
\end{array} \right) = \left( \begin{array}{r}
\bv - A\uv - c_{,\uv}^{T} \sigt \\ 
\lambdav
\end{array} \right),
\end{equation*}
\begin{align*}
    \Lambda &:= C_{,\St} H^{-1} C_{,\St}^T + C_{,\Rt} E^{-1} C_{,\Rt}^T, \\
    \lambdav &:= -\cv + C_{,\St} H^{-1} \left(\psiv + C_{,\St}^T \sigt \right)  + C_{,\Rt} E^{-1} C_{,\Rt}^T \sigt.
\end{align*}

Finally, $\Lambda$ is also block diagonal and positive definite; it can thus also be efficiently inverted, allowing to eliminate $\delta \sigt$ and assemble the Schur complement system
\begin{equation}
    \label{eq:schur_sys}
\left[
     A +  C_{,\uv}^T \Lambda^{-1} C_{,\uv}
] \right] \delta \uv = 
\bv - A\uv + C_{,\uv}^{T} \Lambda^{-1} \lambdav.
\end{equation}

We proceed to solve the symmetric positive semi-definite system~(\ref{eq:schur_sys}) using a Jacobi-preconditioned Conjugate Gradient solver.

\subsection{Performance considerations}
\label{sec:mfem_perf}

 \begin{figure}
     \centering
     \includegraphics[width=\linewidth]{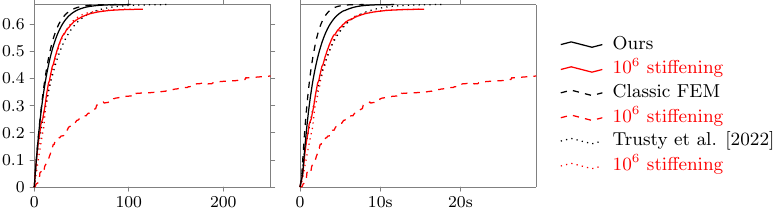}
     \caption{Mean displacement of the dumbbell from Figure~\ref{fig:mfem_comp} as a function of the Newton iteration number~(left) and wall time~(right) for three FEM variants, using an homogeneous material~(black) or with a $10^6$ stiffer center region~(red).
     }
     \label{fig:notched-conv}
 \end{figure}

Overall, we find the per-iteration cost of displacement-only and Mixed FEM to be roughly similar. 
For most of our examples, the cost is dominated by the linear system solve, where both methods share similar unknowns (the per-node displacements) and sparsity stencil after the double condensation step. The assembly of the linear system of Mixed FEM requires multiple sparse matrix--matrix products, but most of those are block--diagonal and cheap to assemble: thanks to our choice of quadrature formula the tensor shape functions are only evaluated at their nodes, where they are Kronecker--delta valued. Conversely, integrating the elasticity Hessian bilinear form for classic FEM requires general interpolation and couples all displacement nodes over a $2^3$ voxels stencil.

One particularity of our method, in contrast to the Mixed FEM technique from~\citet{Trusty22}, is that we explicitly track the rotation $\Rt$ in a separate field. As such we do not need to perform a polar decomposition to extract the symmetric part $\St$ of the deformation gradient when evaluating the constraint residual. While as noted by \citet{Smith18}, this decomposition also gives access to the principal strain basis, which is required to perform the analytical positive semi-definite projection of the elasticity Hessian, in our case the symmetric strain tensor $\St$ is already known, so a $3\times 3$ symmetric eigen-decomposition is sufficient -- and only needs to be done once per Newton iteration rather that for each tentative state in the linesearch as in the approach of \citet{Trusty22}. Moreover, we observe in practice that the convergence of our Mixed FEM solver is not significantly affected by more conservative semi-definite approximations;   the eigenvalues are going to be shifted by the constraint penalization term $\varepsilon$ anyway. In particular, we find that simply ensuring positivity by scaling the volume Hessian term from Neo-Hookean elasticity models according to its minimum eigenvalue, estimated without building the strain basis as per \cite[equation~(30)]{Smith18}, works well.

Another difference is that \citet{Trusty22} express the constraint by placing the nonlinearity on the other side, i.e, as $\Rt^T \Ft = \St$ rather than $\Ft = \Rt \St$. This leads to the derivative form $c_{,\uv}$ being dependent on $\Rt$ and thus $C_{,\uv}$ needing to be re-assembled at each iteration, which can be costly for high-order polynomials; we only need to do this assembly once in our approach. Conversely, we need to reassemble $C_{\St}$ and $C_{\Rt}$ at each iteration, but those, being block-diagonal, are much cheaper to compute. 

While the technique described by \citet{Trusty22} is limited to linear stresses with piecewise-constant rotations and strains, for comparison purposes we have implemented a direct generalization of their method to hexahedral elements using the same choice of basis functions described in Appendix~\ref{app:discrete_mfem}, and report results in Figure~\ref{fig:notched-conv}. For the high-resolution homogeneous dumbbell example, we see that classic FEM is slightly faster than our Mixed FEM, both in terms of iteration count and wall time. In turn, on this example our Mixed FEM is slighly faster than our adaption of the variant from~\citet{Trusty22}. Note we have implemented the three variants in our GPU-based framework, without specific optimizations; results are close enough that different orderings could be observed with other implementations. For the dumbbell with highly stiff center however, we clearly see the displacement-only FEM lagging behind the mixed techniques.

Finally, in contrast with displacement-only and the technique from \citet{Trusty22}, our approach is not fully parameter--free; we need at least the penalization parameter $\varepsilon$ to construct the reduced linear system~(\ref{eq:schur_sys}). Indeed, developing a penalization-free double-condensation step for our approach is hindered by the fact that for a non-isotropic strain $\St$, the matrices $C_{,\St}$ and $C_{,\Rt}$ are not orthogonal, so that $\Lambda^{-1}$ cannot be easily constructed from their pseudo-inverses. A potential solution would be to derive an analytical eigensystem for $\Lambda$, but we reserve this investigation for future work.
While a too strong regularization would slow down convergence, we did not observe this behavior when using our suggested heuristic and setting $\varepsilon$ equal to the typical stress; we see in Figure~\ref{fig:notched-conv} that the convergence curves for our Mixed FEM approach match the non-penalized techniques. Empirically we also obverse that the additional regularization terms help stabilize simulations when the Newton loop is aggressively truncated, as for the interactive clay example depicted in Figure~\ref{fig:clay}.

\end{document}